\documentclass[conference]{IEEEtran}
\IEEEoverridecommandlockouts

\usepackage{graphicx}
\usepackage{textcomp}
\usepackage{xcolor}
\usepackage{hyperref}
\usepackage{booktabs}
\usepackage{lipsum}
\usepackage{comment}
\usepackage{algorithm, algorithmic}
\usepackage{dependencies}
\usepackage{amsfonts}
\usepackage{amsmath}

\usepackage{tikz}
\newcommand*\emptycirc[1][1ex]{\tikz\draw (0,0) circle (#1);} 
\newcommand*\halfcirc[1][1ex]{%
  \begin{tikzpicture}
  \draw[fill] (0,0)-- (90:#1) arc (90:270:#1) -- cycle ;
  \draw (0,0) circle (#1);
  \end{tikzpicture}}
\newcommand*\fullcirc[1][1ex]{\tikz\fill (0,0) circle (#1);}

\begin{document}

\title{SoK: Decentralized Finance (DeFi) - \\Fundamentals, Taxonomy and Risks}

\author{
\IEEEauthorblockN{Krzysztof Gogol\IEEEauthorrefmark{1}, Christian Killer\IEEEauthorrefmark{1}, Malte Schlosser\IEEEauthorrefmark{2}, Thomas Bocek\IEEEauthorrefmark{1}, Burkhard Stiller\IEEEauthorrefmark{1}, Claudio Tessone\IEEEauthorrefmark{1}}
\IEEEauthorblockA{\IEEEauthorrefmark{1}\textit{Computer Science Department, University of Zurich, Switzerland}\\
\IEEEauthorrefmark{3}\textit{Banking and Finance Department, University of Zurich, Switzerland}}
}

\maketitle

\begin{abstract}
Decentralized Finance (DeFi) refers to financial services that are not necessarily related to crypto-currencies. By employing blockchain for security and integrity, DeFi creates new possibilities that attract retail and institution users, including central banks. Given its novel applications and sophisticated designs, the distinction between DeFi services and understanding the risk involved is often complex.
This work systematically presents the major categories of DeFi protocols that cover over 90\% of total value locked (TVL) in DeFi. It establishes a structured methodology to differentiate between DeFi protocols based on their design and architecture. Every DeFi protocol is classified into one of three groups: liquidity pools, pegged and synthetic tokens, and aggregator protocols, followed by risk analysis. In particular, we classify stablecoins, liquid staking tokens, and bridged (wrapped) assets as pegged tokens resembling similar risks. The full risk exposure of DeFi users is derived not only from the DeFi protocol design but also from how it is used and with which tokens. 
\end{abstract}


\section{Introduction}

In 2008 the Global Financial Crisis put the financial markets in turmoil. In the same year, a pseudonym \emph{Satoshi Nakamoto}, an anonymous author or a group of authors, proposed a new decentralized payments system and called it - Bitcoin\cite{Nakamoto2008Bitcoin:System}. Unlike the Traditional Financial system (TradFi), Bitcoin does not depend on any centralized and legal entity, such as a bank acting as a financial intermediary. 
Sparked by Bitcoin, a myriad of new Blockchain (BC) projects and cryptocurrencies emerged, and the total market cap of Bitcoin reached over USD 1 trillion in 2021~\cite{2022Statista}. With the launch of the Ethereum BC~\cite{Buterin2014Ethereum:Platform.} in 2015, a new generation of BCs emerged. In contrast to Bitcoin and its forks, those BCs allowed for the execution of Turing complete Smart Contacts (SC), executed within the virtual machine hosted by the decentralized BC network.
To further advance decentralization, the concept of Decentralized Exchange (DEX) with Automated Market Maker (AMM) emerged in 2016~\cite{ButerinV2016LetsMarkets}. Unlike traditional stock exchanges, the AMM-based DEX does not rely on a centralized order book. This idea was successfully implemented in 2017 by Uniswap\cite{Adams2021UniswapCore}, the first DEX. MakerDAO\cite{MakerDAOTeam2017TheSystem} and Compound\cite{RobertLeshner2019Compound:Protocol} followed with the first decentralized stablecoin and decentralized lending protocol, respectively. In 2016 perpetuals were introduced by BitMEX\cite{Alexander2019BitMEXEffectiveness}, however, still relying on intermediaries. Perpetuals resemble traditional futures contracts without a fixed expiration date to enable derivatives markets for illiquid assets. 
\color{black}
Perpetual without relying on intermediaries emerged with dYdX\cite{Juliano2017DYdX:Derivatives} based on CLOB and, with MCDEX\cite{2023MCDEX}, the Perpetual Protocol\cite{2023PerpetualProtocol}, or GMX\cite{2023GMX} based on AMMs and collateralized vaults in mid 2020.
\color{black}


DeFi, as enabled by public and permissionless distributed ledger, differs from TradFi in various dimensions. First, \1 the \emph{lack of intermediaries} allow parties in DeFi protocols to interact with each other instantly via SCs deployed on BCs. Consequently, the transaction settlement occurs within seconds, 24/7, contrary to days in TradeFi, and is more cost-efficient.  Second, \2 the \emph{permissionless} nature means that any counter-party can interact with the SC, deploy its own SC, and all transactions and SCs are publicly verifiable. For example, any BC reserves or treasuries of DeFi protocols can be instantly audited. Third \3, the \emph{immutability} of the SCs and the BC imply that \emph{code is law} and that nothing can be changed or altered. It should be noted that certain protocols have upgradable SCs, where a certain amount of people are required to perform updates, or Decentralized Autonomous Organizations (DAOs) are voting on specific adaptions. Consequently, DeFi increases efficiency, transparency and accessibility of the financial services\cite{Schar2021DecentralizedMarkets}. Furthermore, the source code of DeFi protocols is often open-source and the governance is executed via DAOs, allowing arbitrary stakeholders to participate in the decision-making process\cite{Jensen2021HOWFINANCE}.

\begin{figure}[h]
\centerline{\includegraphics[width=0.9\linewidth]{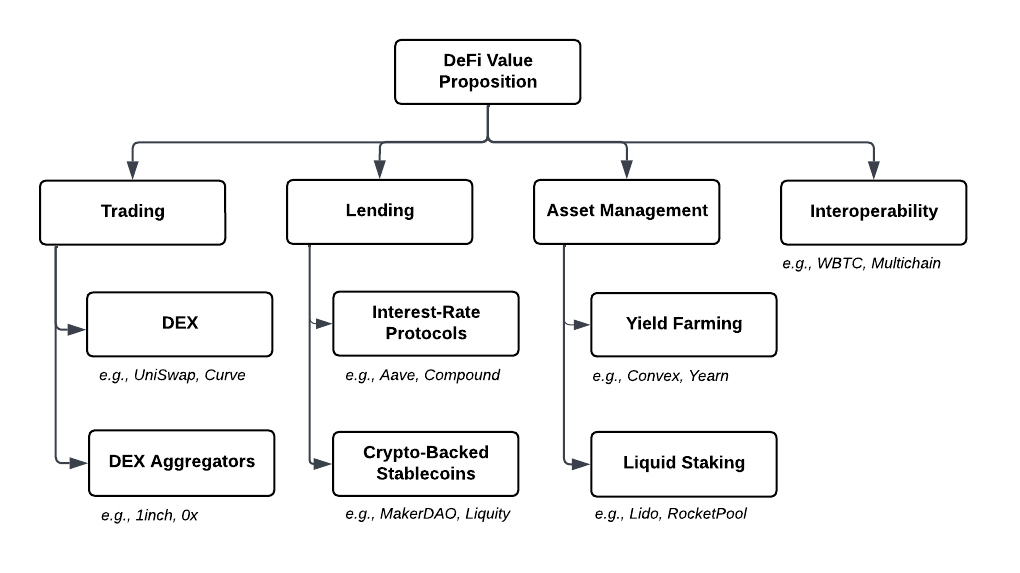}}
\caption{Value Proposition and Design of DeFi Protocols}
\label{fig:DeFiValue}
\end{figure}

With general-purpose and programmable BCs (\eg Ethereum\cite{Buterin2014Ethereum:Platform.}, Solana\cite{2022Solana}, BSC\cite{2022BinanceChain}), it became possible to provide advanced financial services and products without relying on intermediaries for operations, settlement or execution. DeFi applications, called DeFi protocols, allow swapping, lending, margin trading, and borrowing tokens directly within the BC, as depicted in figure \ref{fig:DeFiValue}. Tokens, referred to as digital assets, can cryptographically prove ownership of the real-world asset (\eg art, real estate, and financial products), be linked to other tokens or exist purely digitally for governance, utility, social or other purposes. 
Another popular process is yield farming. It describes the liquidity provisions to DeFi protocols to enable its operations in exchange for participation in the charged fees and thus serves a similar role as market-making in TradFi. The Total Value Locked (TVL), a value representing the value of digital assets locked in yield farming or other DeFi protocols as collateral, grew from zero to USD 200 billion in just two years\cite{2022DeFiLlama} and over five million BC wallet addresses interacted with DeFi protocols\cite{2022StatistaUsers}.



The rapid adoption of the DeFi protocols quickly revealed shortcomings of both protocols and underlying BCs. High transaction costs were the main factor hindering the winder DeFi adoption on the Ethereum BC. Still, Ethereum comprises today 58.5\% of total TVL in DeFi, followed by Tron\cite{2023Tron} (11.0\%) and BSC\cite{2022BinanceChain} (10.6\%) \cite{2022DeFiChains}. Layer 2 BCs on top of Ethereum, \eg Polygon\cite{2023Polygon}, Optimisms\cite{2023Optimism}, zkSynz\cite{2022ZkSync}, starknet\cite{2022Starknet}, emerged as scaling solutions that leverage the security of Ethereum and optimize the transaction costs by taking the calculation off-chain.

\color{black}
\subsection{Methodology and Contribution}

 This work analyses the major, in terms of TVL, categories of DeFi protocols. The initial classification and TVL measures are derived from DeFi Llama\cite{2022DeFiCategories} database. Only DeFi categories, which have TVL above five billion US dollars, are included in the analysis. Those categories represent over 85.71\% of TVL in DeFi as of 24.08.2023 and are listed in the table \ref{tab:DeFiCategories}. DeFiLLama further splits yield farming into yield, aggregator, liquidity manager, and leverage farming subcategories. For the purpose of completeness, we also consider DEX aggregators and algorithmic stablecoins. TVL measure does not apply to DEX aggregators and algorithmic stablecoins. However, DEX aggregators initiate 25\% daily traded volume via DEXs\cite{2022DeFiCategories}. When discussing crypto-backed stablecoins, we introduce and classify all other stablecoins - fiat-backed and algorithmic. Perpetual protocols, with two billion US dollars TVL, are the largest DeFi category that does not meet the indicated threshold. 

\begin{table}[ht]
\centering
\caption{Categories of DeFi protocols\cite{2022DeFiCategories} with Total Value Locked (TVL) denominated in USD}
\begin{tabularx}{\columnwidth}{Xcc} 
    \textbf{DeFi Protocol Category} & \textbf{Protocols} & \textbf{TVL} \\
    \toprule 
    Liquid Staking Tokens & 115 & 20.50b \\
    \hline
    Interest-rate Protocols & 289 & 13.06b \\
    \hline
    Decentralized Exchanges & 975 & 12.20b \\
    \hline
    Bridges and Wrapped Tokens & 47 & 8.85b \\
    \hline
    Crypto-backed Stablecoins & 100 & 8.37b \\
    \hline
    Yield Farming Aggregators & 364 & 5.02b \\
    \hline
     Other & 656 & 7.18b \\
    \hline
    \bottomrule  
    \end{tabularx}
    \label{tab:DeFiCategories}
\end{table}

Thus, this work presents the first Systematization of knowledge (SoK) that \1 applies a data-driven approach for the selection of DeFi protocols' categories, and \2 combines the vertical and horizontal research in order to present a comprehensive overview of DeFi.
It contributes in the following areas:
\begin{itemize}
\item  The main contribution of this SoK is the first risk framework for DeFi that clearly and exhaustively assigns DeFi risk factors to the protocol category, stakeholder, and type of the underlying token. The exposure to DeFi risk can be derived from these three factors.
\item This is the first DeFi SoK that took a data-driven approach to build the taxonomy for DeFi protocols. It separates the independent classifications by design (algorithm) and by value proposition. For example, interest-rate protocols and crypto-based stablecoins are both lending protocols, whereas technically, interest-rate protocols are lending pools and stablecoins are synthetic tokens.
\item It is the first SoK that included liquid staking tokens (LSTs), the largest class of DeFi protocols, and showed that LSTs, due to the same design as stablecoins, are prone to de-peg risk.
\end{itemize}

\subsection{Related Works}
Numerous SoK and surveys on DeFi protocols exist. First, related works mostly provide a high-level overview of DeFi and, second, these works arbitrarily mention selected categories of DeFi protocols\cite{Auer2023TheDeFi, Werner2021SoK:DeFi, Carapella2022DecentralizedRisks, Schar2021DecentralizedMarkets, Jensen2021AnDeFi, Green2022DeFiInstitute, Zutshi2021ThePlatforms}.

\cite{Werner2021SoK:DeFi} shortly presents DEXs, lending, portfolio management portfolio, and stablecoins, focusing on the technical and application layer threats to DeFi, and stakeholders DeFi depends on. The additional sub-classification and mathematical assumptions for AMM-DEX are presented in  \cite{Schar2021DecentralizedMarkets} with a brief introduction to lending and asset management categories. This work extends the previous SoK based on the data-driven approach, especially regarding liquid staking, new AMM types, and risk analysis.

The DeFi layers, stakeholders, and token classification are introduced in \cite{Auer2023TheDeFi} for DEX, lending, and derivatives protocols.
The history of DeFi attacks, followed by the technical classification, is presented in \cite{Zhou2022SoK:Attacks}. This research extends that work with the general framework of threats to DeFi users, token and protocols.

Another group of SoK focuses on the vertical analysis of specific DeFi categories, \eg decentralized exchanges with automated market making\cite{Xu2021SoK:Protocols}, yield farming\cite{Cousaert2021SoK:DeFi, Xu2022ReapProtocols}, lending\cite{Gudgeon2020DeFiEfficiency} or stablecoins\cite{Bullmann2019OccasionalSolution, Moin2019ADesigns}. When appropriate, this work directs the reader to such SoKs for further material.

\color{black}
\section{Background}

The general design principles for blockchains and Distributed Ledger Technology (DLT) and their implications for activities in financial services have been widely studied \cite{Jensen2021AnDeFi}. This section introduces the necessary blockchain definitions that are later used in the paper.

\subsection{Layer-1 and Layer-2 Blockchains}
A blockchain (BC) is a computer network that maintains a distributed database. This network only allows appending new transactions and forbids deleting or updating any transactions. Transactions are grouped into blocks and block order in the database is maintained, and verifiable, using cryptographic hash functions and digital signatures. 
BCs can be classified as permissioned and permissionless. Permissionless BCs are open (public) networks accessible by all. Permissioned BCs can be accessed only by external parties recognized by the system administrator \cite{Labazova2019TowardsBlockchain}. For this paper, we define a BC as a public - trustless and permissionless - distributed ledger. Consequently, DeFi applications are also trustless - do not depend on any centralized entity - and permissionless - open to use by any BC wallet. 

\begin{table*}[t]
\centering
\renewcommand{\arraystretch}{1.2}
\caption{Terminology used in this work}
\begin{tabularx}{\textwidth}{llX} 
    \textbf{Abbreviation}& \textbf{Term}& \textbf{Definition} \\
    \toprule 
    BC  & Blockchain & Public data storage executing a permissionless consensus protocol in a decentralized deployment~\cite{TODO_CK} \\
    \hline
    SC  & Smart Contract & Digital protocol implemented in software, executing the terms of a contract agreed between two or more parties without the need of an intermediary~\cite{TODO_CK} \\
    \hline
    L1 & Layer 1 & BC with its own independent network executing a consensus protocol security (\eg BC with its own set of validators or miners) \\
    \hline
    L2 & Layer 2 & An overlay network that relies on other BC for the security (\eg immutability)  \\
    \hline
    DeFi & Decentralized Finance & Financial applications that rely on BC for their security and integrity (\cf Tab.~\ref{tabDeFiCeFiTradFi}) \\ 
    \hline
    CeFi & Centralized Finance &  Financial applications that are hosted outside of BC, but involve BC in their business model (\cf Tab.~\ref{tabDeFiCeFiTradFi})  \\
    \hline
    TradFi & Traditional Finance & The traditional financial system, unrelated to BC or DeFi (\cf Tab.~\ref{tabDeFiCeFiTradFi}) \\
    \hline
    CEX & Centralized Exchange & Exchange that allows swapping fiat currency to cryptocurrencies and tokens  \\
    \hline
    DEX & Decentralize Exchange & Exchange that operates entirely on the BC and allows to swap cryptocurrencies and tokens \\
    \hline
    AMM & Automated Market Model & Mathematical formula used to determine the exchange rate between tokens\\
    \bottomrule  
    \end{tabularx}
    \label{tab:Definitions}
\end{table*}

A network \emph{consensus mechanism} defines the rules, for what constitutes a legitimate transaction and block. The dominant consensus mechanisms are proof-of-work (PoW) and proof-of-stake (PoS). The consensus mechanism economically incentives network participants - miners in PoW and validators in PoS blockchains - to promote network security. In PoW BCs, miners, to append a new block to the ledger, are required to solve the cryptographic puzzle in a process that requires electricity\cite{Nakamoto2008Bitcoin:System}. In PoS BCs, validators, to append a new block, must pledge ("stake") a certain amount of native tokens of the underlying BC. In exchange for appending a new block, validators receive a reward. In case of malicious behavior, validators are punished, in a process called \emph{slashing} by a fee from the staked tokens. \emph{Staking} is the process of providing BC native tokens to validators to participate in profits from the block rewards. By design, PoS blockchain requires validators and \emph{stakers} to freeze their tokens for a period of time that may vary from days to years. \emph{Gas fee} is paid to miners and validators for appending the transaction to the BC. 

\paragraph{Layer-1 vs Layer-2}
According to \emph{blockchain trilemma}, the blockchain network can only prioritize two features between three: decentralization, security or scalability\cite{Conti2019BlockchainMessages}. As security is an absolute requirement and decentralization is a promise of BC, scalability has remained a challenge, resulting in low  transaction  rates  and  high  transaction  processing  latencies. \emph{Layer-1} is the blockchain with its own, independent trust assumptions: the network of nodes, and consensus mechanism \cite{Gangwal2022AProtocols}. Layer-1 solutions target the improvements of the core elements of blockchain design, \eg block data, consensus mechanisms, or sharding the network. \emph{Layer-2} solutions aim to scale the BC without modifying the underlying trust assumptions. Layer-2 protocols are built on top of layer-1 BCs. \emph{Roll-ups} are the most common scaling solutions for DeFi protocols deployed to Ethereum\cite{2022L2BeatTVL}. They aim to reduce the load of the main chain by taking the transaction execution off the chain in batches and bundling them together for on-chain verification. Depending on the verification process, roll-ups can be divided into two groups: optimistic roll-ups, \eg Arbitrum\cite{2022Arbitrum}, Optimism\cite{2023Optimism}, and zero-knowledge (zk) roll-ups, \eg zkSync\cite{2022ZkSync}, starknet\cite{2022Starknet}. Table \ref{tab:Definitions} summarizes critical definitions and abbreviations used in the paper. 

\subsection{Centralized Finance}
\emph{Decentralized Fiance} (DeFi) is a general term that refers to all financial applications that rely on blockchain for their security and integrity. \emph{Centralized Finance} (CeFi) describes the financial applications that are hosted outside of BC, but involve digital assets in their business. An example of such CeFi application is \emph{centralized exchange} CEX \eg Binance \cite{2022BinanceExchange} or Kraken\cite{2022KrakenExchange}, which allows exchanging the fiat currency (USD, EUR, CHF, etc) into cryptocurrencies. Those cryptocurrencies purchased at CEX can be further transferred to the crypto wallet on BC or remain in the custody of CEX. \emph{Traditional Finance} (TradFi) refers to the traditional finance system that does not involve digital asset-related activities or products.


\begin{table}[ht]
\centering
\caption{Comparison of DeFi, CeFi, and TradFi~\cite{Qin2020AttackingProfit}, \\ \emptycirc = Yes, \fullcirc = No, \halfcirc = depends on the specific protocol}
\begin{tabularx}{0.45\textwidth}{Xcccc} 
    & \makecell{\textbf{BC} \\ \textbf{Related}} & \makecell{\textbf{BC} \\ \textbf{Settlement}} & \makecell{\textbf{Non} \\ \textbf{Custodial}}  & \makecell{\textbf{DAO} \\ \textbf{Governance}} \\
    \toprule 
    DeFi   & \fullcirc     & \fullcirc   & \fullcirc       & \halfcirc      \\ 
    CeFi   & \fullcirc      & \halfcirc   & \emptycirc      & \emptycirc     \\ 
    TradFi & \emptycirc      & \emptycirc  & \emptycirc      & \emptycirc   \\ 
    \bottomrule  
    \end{tabularx}
    \label{tabDeFiCeFiTradFi}
\end{table}

Boundaries between CeFi and DeFi are not always clear and the work \cite{Qin2021CeFiFinance} provides a structured methodology for differentiation. If the user is not in control of the tokens (\eg does not retain custody, cannot transfer the assets without the financial intermediary), the financial application classifies as CeFi. CeFi applications may operate with BC settlement or not. In November 2022, FTX, the second largest CEX in the world collapsed, leaving its customers without access to their cryptocurrencies\cite{2022FTXDossier}. If the user is in control of the digital assets, but the financial intermediary can censor the transaction, such application classifies as CeFi with the BC settlement. Examples include USDT and USDC - stablecoins that are backed by fiat money. The issuer of USDT destroyed 44M of blacklisted USDT\cite{Qin2021CeFiFinance}. DeFi protocols can be further classified by their governance model: centrally governed and DAO. Centrally governed DeFi protocols are governed from outside of the BC. In the DAO model, decisions are made by holders of governance tokens via voting.

\section{DeFi Taxonomy}
DeFi protocols provide various financial services and can be classified based on the value proposition to users: trading, lending, asset management, and BC interoperability, as depicted in table \ref{fig:DeFiValue}. This section proposes novel classification frameworks. First, we separate the token classification from the taxonomy of DeFi protocols. Next, we classify DeFi protocols in two dimensions - algorithm and network (blockchain) architecture.
\emph{Taxonomy of DeFi algorithms} is based on the technical and economical design of DeFi protocols. The DeFi algorithm determines how financial services are performed on the BC and to which risks DeFi agents are exposed.
\emph{Taxonomy of DeFi network architecture}, refers to the number of BCs, on which the protocol operates and the coordination of its work across those BCs. The network architecture implies the infrastructure risks.
The classifications within those two domains, as well as token taxonomy, are straightforward, as the values assigned to each dimension can be traced from the DeFi protocols' implementations and whitepapers that are publicly available.


\subsection{Application}
As summarized in figure \ref{fig:DeFiValue}, DeFi protocols provide various value propositions to users - trading, lending, asset management and BC interoperability - and extend the initial objective of BCs: a decentralized payment system and store of value\cite{Nakamoto2008Bitcoin:System}.

\paragraph{Trading}
\emph{Decentralized Exchanges} (DEXs), \eg UniSwap\cite{Adams2021UniswapCore}, allow DeFi users to swap between two tokens with the settlement occurring directly on the blockchain. Compared to Centralized Exchanges (CEXs), DEXs offer higher diversity of tradeable tokens and higher security of transactions. \emph{DEX Aggregators}, \eg 0x\cite{Warren20170x:Blockchain}, are DeFi protocols that collect exchange rates from various DEXs and the best execution options. The major risks of DEXs for DeFi users include slippage risk and MEV attacks. Slippage cost refers to the difference between the price quoted by the DEX and the executed price. The MEV (maximum extractable values) attacks are executed by miners or validators that re-order transactions for their profits, leaving the DeFi user with higher slippage costs. 

\paragraph{Lending}
Lending Protocols allow DeFi users to borrow tokens against collateral. \emph{Interest rate protocols}, such as Compound\cite{RobertLeshner2019Compound:Protocol}, and Aave\cite{2020AaveV1.0}, create pools of tokens that can be lent and borrowed, thus they are also referred to as protocols for loanable funds. Depending on token supply and demand, the protocols automatically adjust interest rates. Another mechanism for on-chain lending is provided by \emph{crypto-backed stablecoin} protocols, such as MakerDAO\cite{MakerDAOTeam2017TheSystem}, and Liquity\cite{2023Liquity}. Stablecoins are tokens that hold their value pegged to the fiat currency, typically US dollar. In order to hold the peg to the reference value, crypto-backed stablecoin holds collateral of other crypto tokens to ensure that the circulating token has a redemption value. Any DeFi user can borrow stablecoins directly from the stablecoin protocols by proving crypto collateral. In such a case, a collateralized debt position (CDP) is created. The major risk of DeFi lending is the liquidation risk of the underlying collateral. The liquidation occurs when the collateral value descends behind the minimum over-collateralization threshold. The dep-peg risk refers to the circumstances when stablecoin fails to hold the reference value.

\paragraph{Asset Management}
The DeFi Protocols in asset management allow DeFi users to generate additional yield on their tokens. \emph{Yield farming} is a process of providing liquidity capital to DeFi protocols in exchange for fee participation rewards\cite{Xu2022ReapProtocols}\cite{Xu2021SoK:Protocols}, similarly to market making in traditional finance. \emph{Impermanent Loss} is a major economic risk associated with yield farming, caused by the high volatility of the crypto-tokens. \emph{Liquid staking} refers to DeFi protocols that accumulate rewards from staking without locking the tokens. The staking process at PoS BCs supports the security of the underlying BC, but requires locking staked tokens for a defined period to collect the staking reward. \emph{De-Peg risks} is the major risk related to liquid staking. 

\paragraph{Blockchain Interoperability}
Blockchains, by design, are siloed computer networks that do not communicate with each other. Bridges allow DeFi users to \1 transfer tokens between blockchains without employing any CEX, and \2 and to hold exposure to the tokens from other blockchains, \eg exposure to Bitcoin on the Ethereum chain. To achieve this objective, bridge protocols create \emph{wrapped tokens} - tokens with the value pegged to the tokens at other blockchains, \eg wrapped Bitcoin\cite{2023WrappedBitcoins} on the Ethereum chain. Wrapped tokens are exposed to the de-peg risk. 

\subsection{Design}

\begin{figure*}[h]
\centerline{\includegraphics[width=\linewidth]{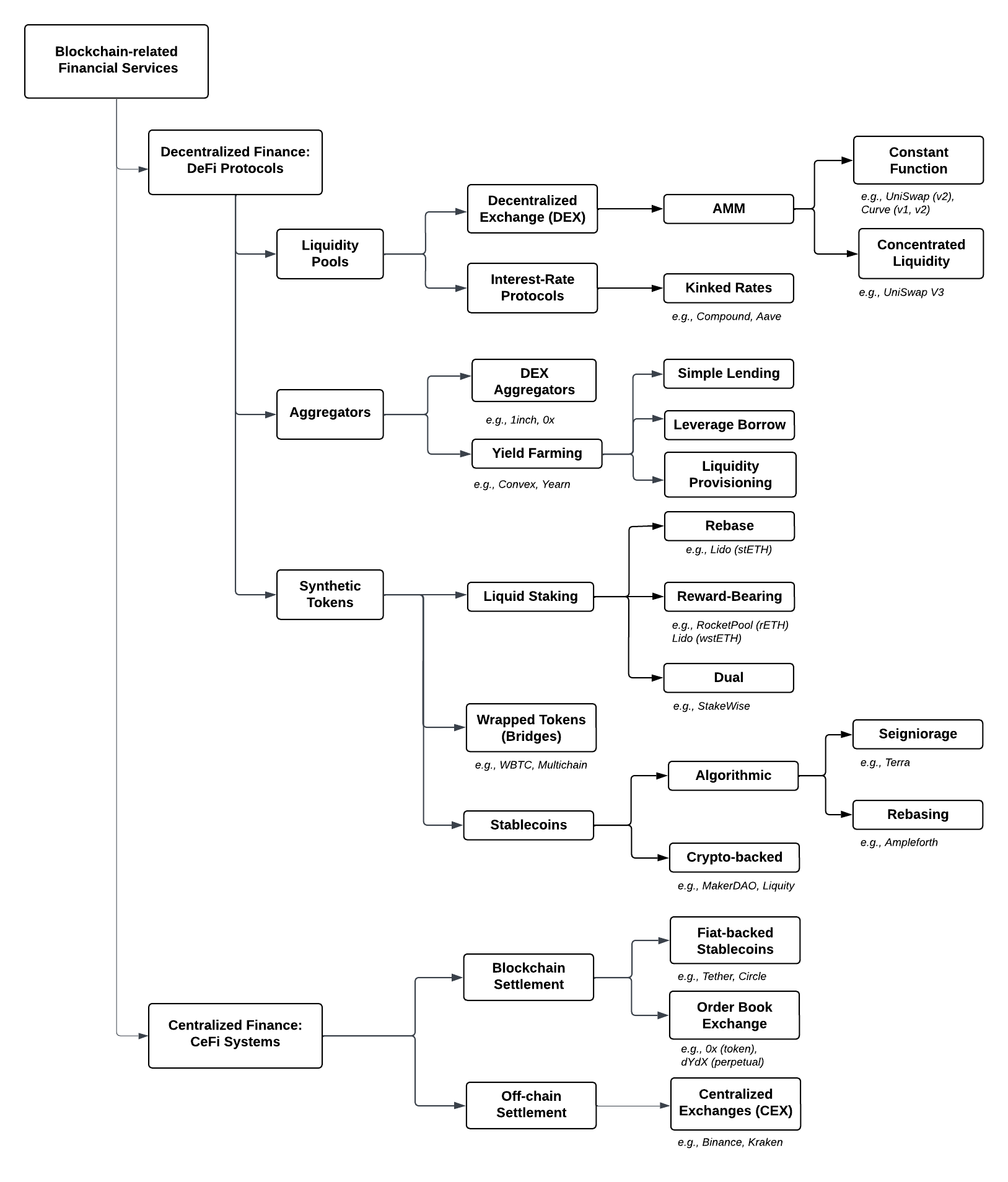}}
\caption{Taxonomy of blockchain-based financial application}
\label{fig:taxonomy}
\end{figure*}


In the design domain, the DeFi protocol can be classified into three dimensions: \1 liquidity pools \2 aggregators \3 synthetic tokens, as presented in figure \ref{fig:taxonomy}. The DeFi protocols are developed on the infrastructure layer of their underlying blockchains. The infrastructure layer includes native tokens or staking for proof-of-stake blockchains.
\paragraph{Synthetic (Pegged) Tokens} Synthetic tokens are similar to derivatives in traditional finance. They employ smart contracts to keep the peg to the target value, which might be fiat currency - stablecoins, tokens at other blockchains - wrapped tokens of bridges, and staked tokens - liquid staking. The inherent risk of all synthetic tokens is the re-peg risk. 

\paragraph{Liquidity Pools} Liquidity pool-based DeFi protocols match the demand and supply using token pools operated by smart contracts. The major DeFi liquidity pool-based protocols include decentralized exchanges (DEX) and interest-rate protocols. 
DEXs manage the supply and demand to swap tokens and, depending on the design, can be further classified as CLOB or AMM. There are various implementations of AMM, with concentrated liquidity being the latest improvement that increases the capital efficiency at AMM DEX. 
The interest-rate protocol manages the supply and demand for borrowing and lending and adjusts interest rates automatically. DeFi liquidity pools protocols and stablecoins are referred to as DeFi primitives \cite{Cousaert2022SoK:DeFi}. 

\paragraph{Aggregators} The major aggregator protocols facilitate the operations of DEXs and include DEX aggregators and yield farming protocols. DEX aggregators allow DEX service customers - traders - to find the best execution rate, and sometimes offer protections again MEV attacks, by splitting the transaction into smaller ones. Yield farming protocols simplify and increase liquidity efficiency at AMM DEXs.  

\subsection{Architecture}
This dimension indicated whether DeFi protocols operate on a single chain or across multiple chains. Protocols operating on multiple networks are in the early stage but they are expected to play a substantial role in DeFi in the future\cite{Freni2022TokenomicsFramework}.

\paragraph{Single Chain} Single-chain protocols are deployed and operate within one BC. Most of DeFi protocols operate on single chain\cite{2022DeFiLlama}.
\paragraph{Cross Chain} Cross-chain protocols that operate on multiple BCs and allow the value to cross between BCs. Those protocol utilize blockchain interoperability technologies so that the value stored on one BC is transformed into value at other BC\cite{Lin2021OverviewTechnology}. 
The cross-chain protocols include bridges.
\paragraph{Multi Chain} Multi chain protocols are deployed to multiple blockchain, at which they operate independently. Contrary to cross-chain protocols, they do not utilize value cross BC, \eg cross-chain collateral. Example of DeFi protocols deployed to multiple chains include Aave\cite{2020AaveV1.0}, Uniswap\cite{Adams2021UniswapCore}.
\paragraph{App Chain} App-chain (or Solo-chain) is DeFi protocol that developed sovereign BC to optimise its services for the specific use cases. Depending on the security model of the sovereign BC, the app-chain can be classified at L1, L2 or L3. Cosmos\cite{2022Cosmos} provides SDK for the development of L1 appchain. Polkadot\cite{2022Polakdot} simplifies the development of L2 appchains, called parachains by Polkadot. Starknet\cite{2022Starknet} and zkSync\cite{2022ZkSync} are zero-knowlege rollups (L2) on top of Ethereum and offer possibilities to built furhter L3.


\begin{figure*}[h]
\centerline{\includegraphics[width=0.8\linewidth]{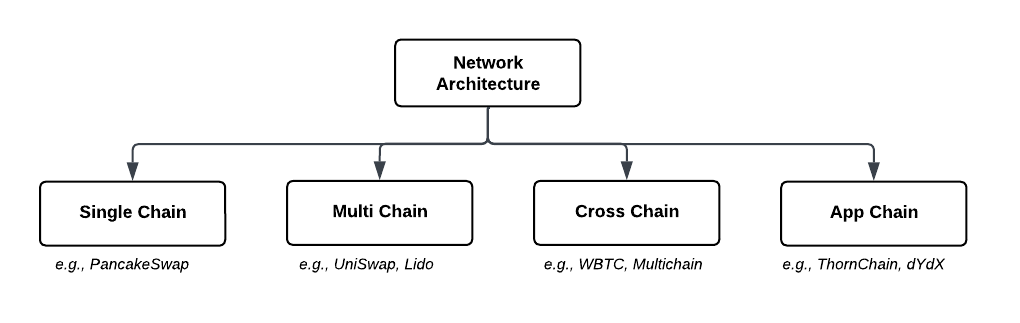}}
\caption{Taxonomy of DeFi Network Architecture}
\label{fig:taxonomy}
\end{figure*}

\subsection{Governance}
DAO vs Centralized

\subsection{Related Classifications}

\subsubsection{DeFi Stakeholders}
Blockchain is a decentralized network, in which its stakeholders (nodes, validators) operate in \emph{peer-to-peer model}, without any system administrators or centralized entity. DeFi protocols are smart contracts deployed to BC and DeFi users interact with those smart contracts. \emph{Liquidity pool} is a smart contract that holds tokens deposited in a DeFi protocol by \emph{liquidity providers} (LPs) \cite{Inzirillo2022ManagingPortfolios} in order to facilitate financial services (\eg borrowing, token swapping). LPs thus fulfill a similar role to that of market makers in TradFi. DeFi stakeholders - service customers of DeFi protocols, LPs, arbitrageurs, and governance users (\eg founders, developers, investors) interact with the liquidity pools, operating in \emph{peer-to-pool} model\cite{Jensen2021AnDeFi, Auer2023TheDeFi}. DeFi stakeholders have different incentives and roles in DeFi, and consequently are exposed to various risks. The table \ref{tab:DeFiAgents} summarizes the DeFi stakeholders and risk analysis follows in the further sections. 

\begin{table}[t]
\centering
\caption{DeFi stakeholders, based on~\cite{Jensen2021AnDeFi, Auer2023TheDeFi}} 
\begin{tabularx}{0.9\columnwidth}{XXX} 
    \textbf{Stakeholder} & \textbf{Role} & \textbf{Incentive} \\
    \toprule 
    \textbf{Service} \textbf{Customers}
    & Interacts with the DeFi protocol & Profit, credit, liquidity mining \\
    \hline
    \textbf{Liquidity} \textbf{Providers} & Provides capital to DeFi protocols to ensure sufficient liquidity for financial services & Protocol fee profit participation, liquidity mining \\
    \hline
    \textbf{Arbitrageurs} 
    & Eliminate the market inefficiencies between different DeFi protocols & Arbitrage Profits \\
    \hline
    \textbf{Governance} \textbf{Users} & Design, develop and maintain the DeFi protocol & Governance token appreciation \\
    \bottomrule  
    \end{tabularx}
    \label{tab:DeFiAgents}
\end{table}

\emph{Liquidity mining} is a process of interacting with a DeFi protocol to collect its governance tokens or other rewards. The process is not related to mining in PoW BCs\cite{Inzirillo2022ManagingPortfolios}. The prime objective of liquidity mining is to attract new users and LPs to DeFi protocols.


\subsubsection{Token Taxonomy}
There exist multiple extensive token classification frameworks \eg \cite{Freni2022TokenomicsFramework, Ankenbrand2020ProposalTaxonomy, Auer2023TheDeFi}. Based on the work of \cite{Jensen2021AnDeFi} and \cite{Rahman2022SystematizationManagement} we select the two token classification domains: "technology" and "underlying value". Token technology comprises two dimensions: BC native tokens (\eg Bitcoin, Ethereum, Solana) and smart-contract-enabled tokens. The "underlying value" dimension comprises three dimensions: asset-backed, network value, and share-like\cite{Freni2022TokenomicsFramework}. Asset-backed tokens comprise tokens linked to other assets (\eg tokens confirming participation in liquidity pools), derivative tokens with other tokens as underlying (\eg stablecoins, liquid staking tokens), and real-world assets - tokenized assets from the real world.
The proposed taxonomy, which combines and simplifies, other classification frameworks \cite{Freni2022TokenomicsFramework}, emphasizes the risk of the token holder. This token classification can help to ensure a consistent view of the different token types and the underlying counterparty risk.

\begin{figure*}[h]
\centerline{\includegraphics[width=0.9\linewidth]{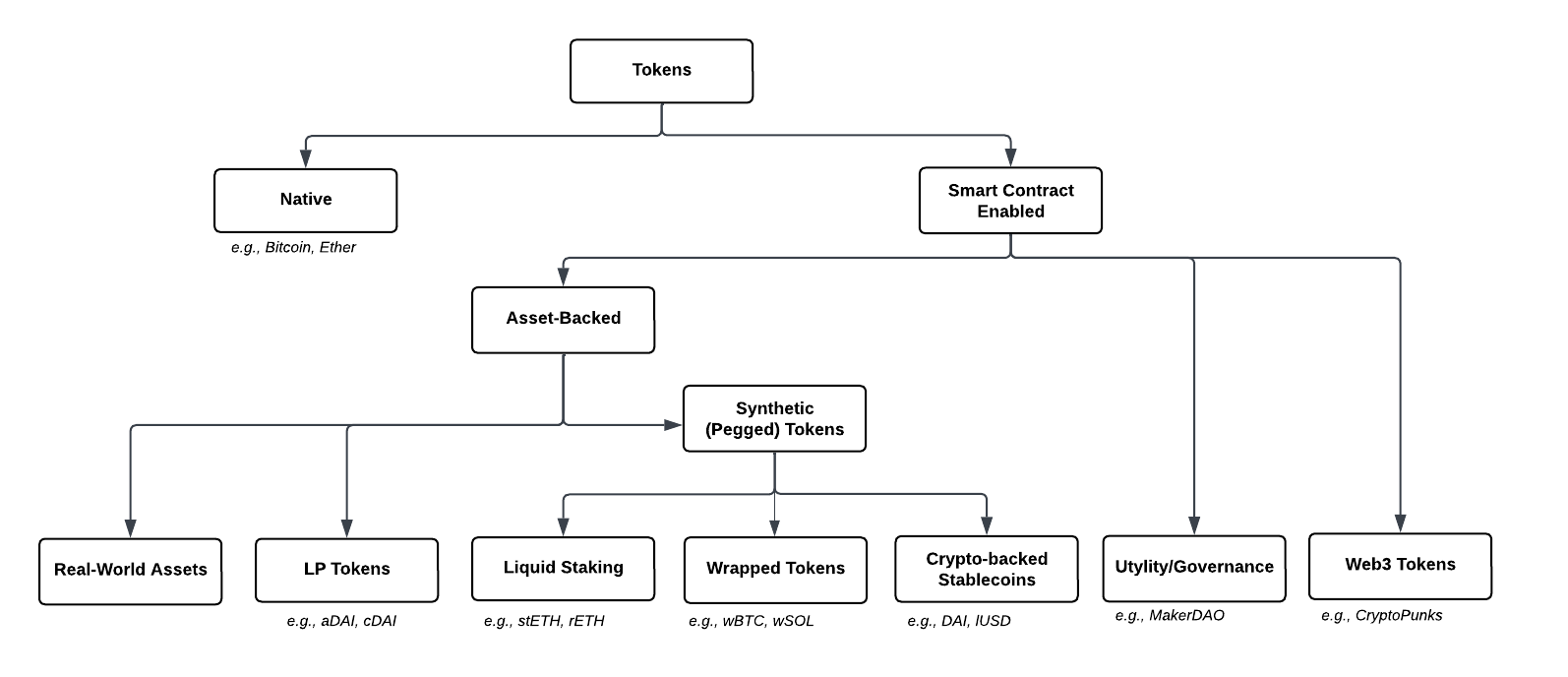}}
\caption{Taxonomy of Tokens}
\label{fig:taxonomy}
\end{figure*}

\begin{itemize}
    \item \textbf{Native tokens}: Layer 1 BC's native tokens, \eg Bitcoin, and Ethereum. 
    \item \textbf{Smart contract enabled tokens}: Tokens that are created within BC by smart contracts can be further divided into more granular dimensions:
    \begin{itemize}
        \item \textbf{Utylity/Governance (Network Value)}: the underlying value is represented by the level of trust, \eg DAO governance tokens or other utility tokens for fee payments within the smart contract applications. Network-driven demand for the token may lead to an increase in value in case of limited supply.
        \item \textbf{Asset-backed} 
            \begin{itemize}
                \item \textbf{LP Tokens (On-chain Asset)}: tokens representing ownership of other on-chain assets, \eg  tokens confirming participation in the liquidity pools of AMM-DEX or interest-rate protocols,
                \item \textbf{Synthetic (Pegged) Assets}: tokens pegged to the target value, \eg crypto-backed stablecoin, liquid staking token, or wrapped Bitcoin (for bridged tokens). The DeFi protocol minting and burning the synthetic token is responsible for maintaining the peg.
                \item \textbf{Real-World Assets}: the underlying value is tied to the tokenized real-world assets, \eg fiat money for fiat-backed stablecoins.
            \end{itemize}
        \item \textbf{Web3 (Share-like) Tokens}: the underlying value is generated by the work of the token holder or 3rd party issuers \eg NFT representing art ownership.
    \end{itemize}
\end{itemize}

\section{DeFi Protocols} \label{DeFiProtocolFundamentals}
This section presents the mechanics behind the major DeFi categories: decentralized exchanges (DEXs) and their aggregators, lending protocols (also referred to as interest-rate protocols or protocols for loanable funds), crypto-backed stablecoins (also referred to as collateralized debt positions) with other classes of stablecoins, bridges (wrapped tokens) and liquid staking protocol (liquid staking tokens/derivatives).

\subsection{Pegged Tokens}

\subsubsection{Liquid Staking Tokens}

\emph{Synthetic assets} in DeFi are analogous to derivatives in traditional finance (TradFi) \cite{Rahman2022SystematizationManagement}. In TradFi, derivatives allow traders to profit from price movements of stocks and bonds they do not own\cite{FernandoJ2022Derivatives:Cons}. Synthetic assets are cryptocurrency tokens that track the price fluctuations of the target asset\cite{Rahman2022SystematizationManagement, BajpaiP2022CryptocurrencyFuture}. \emph{Liquid staking} is a synthetic asset that tracks the value of the tokens staked in the proof-of-stake blockchains \cite{Scharnowski2022LiquidDiscovery}. Liquid staking token is also referred to as \emph{liquid staking derivative} (LSD) and the protocol minting it - \emph{liquid staking protocol} (LSP).
Staking is the process of locking tokens in validators for future profit \cite{Scharnowski2022LiquidDiscovery}. Validators, network participants in PoS blockchain, put those tokens at stake when appending new transactions to the blockchain. For each correctly appended transaction, the validator receives a reward, which is redistributed between token providers (\emph{stakers}). The dishonest behavior is punished by a reduction in the staked tokens (\emph{slashing}). By design, the staked tokens are locked for a specific period, making it impossible to use them in other applications, reducing the token liquidity. 
By tracking the value of the staked assets, liquid staking tokens allow participating in the staking benefits. Unlike the staked tokens, liquid staking tokens are not locked but are freely traded and used in other DeFi applications. There are three standard LSD models \cite{Scharnowski2022LiquidDiscovery, 2023AProtocols:}: \1 rebasing tokens, \2 reward-bearing tokens, and \3 dual-token model.

\paragraph{Rebase Tokens} In this model, the LSP tokens keep the 1:1 peg to the staked tokens and inflate the supply to reflect the yield earned from staking. The most popular rebase LSD is stETH, minted by Lido Protocol\cite{2020Lido:Whitepaper}. Lido every day increases the balance of all stETH holders to reflect the additional ETH earned as staking rewards. The advantage of rebasing LSP is its simple functionality. The disadvantage is lack of the compatibility with other DeFi protocols, \eg DEXs or interest rate protocols. 

\paragraph{Reward-bearing Token} In this model, the token does not maintain 1:1 peg to the staked native token, but increases its value to reflect the staking yields. Consequently, the reward-bearing token grows in value, related to the underlying native token. Reward-bearing tokens provide better compatibility with DeFi protocols. Most of LSPs follow this design: RocketPool\cite{2023RocketPool}, Marinade \cite{2023Marinade}, CoinBase Wrapped Staking \cite{2022CoinBaseWhitepaper}.

\paragraph{Dual Token} In the dual token model the asset token is split from the staking revenue it yields. For example, StakeWise \cite{2023StakeWise} operates in the two token model: sETH2 - with 1:1 peg to ETH -  and rETH2 - that reflects the earned staking rewards. The disadvantage of this model is the fragmented liquidity between two tokens.

\begin{table}[t]
\centering
\caption{Overview of selected liquid staking tokens}
\begin{tabularx}{0.8\columnwidth}{XXXX} 
    \textbf{Tokens} & \textbf{Protocol} & \textbf{Blockchain} & \textbf{Model} \\
    \toprule 
     stETH & Lido & Ethereum & rebasing \\
    \hline
    wstETH & Lido & Ethereum & reward-bearing \\
    \hline
    mSol & Marinade & Solana & reward-bearing \\
    \hline
    rETH & Rocket Pool & Ethereum & reward-bearing \\
    \hline
    sETH2, rETH2 & StakeWise & Ethereum & 2 tokens \\
    \bottomrule  
    \end{tabularx}
\label{tab:LiquidStakingTokens}
\end{table}

\paragraph{Other Classifications} LSP also varies in the validator selection process: \1 whitelisting reputable validators, \2 credentials-based, \3 collateral. In the whitelisting model, LSP adds only trusted validators to the pool of validators, \eg Lido\cite{2020Lido:Whitepaper}. In the credential based-model, the pool of validators is permissionless upon fulling the minimum criteria. In the last model, LSP requires validators to post the collateral to guarantee its performance, \eg Rocket Pool\cite{2023RocketPool}. The collateral model hampers the growth of the validator pools by increasing the minimum threshold for staking. 

\subsubsection{Bridges and Wrapped Tokens}
DeFi ecosystem has evolved into a multi-chain ecosystem, but each BC is an independent and closed network by design. Miners and validators write transactions to the BC according to its consensus mechanism. Fragmentation of DeFi protocol and liquidity across various chains hinders DeFi growth and locks DeFi users to a single BC. Blockchain interoperability includes the transfer of digital assets, data, and smart contract instructions between BCs, which would allow DeFi users to utilize digital assets stored in one BC in DeFi protocols on another BC (\emph{cross-chain collateral}) and execute low-cost and fast transaction of digital assets stored on less scalable BC (\emph{efficiency and scalability}). For the DeFi protocols, blockchain interoperability would allow for more efficient utilization of liquidity, locked across various BCs \cite{2021WhatDeFi}.
While Polkadot\cite{2022Polakdot} and Cosmos\cite{2022Cosmos} provide frameworks to develop interoperable BCs, \eg by sharding, they do not facilitate cross-chain communication with networks outside of their BC-framework\cite{Zamyatin2019XCLAIM:Assets}. Centralized exchanges (CEX) can be used to transfer digital assets across blockchains. However, given their centralized nature, they do not deliver on BC and DeFi promises of decentralization, permissionless and trustless. 
 
\emph{Bridges} enable the transfer of tokens between BCs  \cite{Haque2022SemanticSummarization, Mccorry2021SoK:Blockchains, Hardjono2021BlockchainHash-Locks, Wang2021SoK:Interoperability, Lafourcade2020AboutInteroperability, Stone2021TrustlessBridges, Caldarelli2022WrappingTokens, Lee2022SoK:Hacks, Zarick2021Delta:Trilemma}, \eg Bitcoin and Ethereum, or between layers of the same BC, \eg Ethereum and its roll-ups. Based on the underlying architecture, bridges can be classified as \1 wrapped tokens, \2 liquidity networks and \3 hybrid solutions\cite{Zamyatin2019XCLAIM:Assets}. 




\subsubsection{Stablecoins}
Stablecoins are tokens, which value is pegged to a fiat currency, typically 1 USD. They provide a way for DeFi users to mitigate the volatility of other digital assets, without the need to transfer liquidity outside the blockchain\cite{Moin2019ADesigns}. Holders of stablecoins are exposed to the \emph{de-peg risk} - the risk that the stablecoin will not hold the peg to the target value. Stablecoin protocols employ various mechanisms to hold the peg and depending on the collateral type, they can be classified into three categories: \1 fiat-backed, \2 crypto-backed, and \3 algorithmic stablecoins\cite{Bullmann2019OccasionalSolution}. The collateral ensures that the circulating token has redemption value. 
\paragraph{Crypto-backed stablecoins}{Fiat-backed stablecoins}
Fiat-backed stablecoins utilize collateral of a fiat currency or commodity outside the blockchain, thus, they are also referred to as off-chain collateralized stablecoins. Given the collateral held outside of the blockchain and regulatory restrictions, fiat-backed stablecoins are not considered to be part of DeFi, but CeFi\cite{Qin2021CeFiFinance}.
\paragraph{Crypto-backed stablecoins}
Crypto-backed stablecoins maintain the collateral of other crypto tokens on the blockchain. They are also referred to as on-chain collateralized or over-collateralized stablecoins. Over-collateralization implies that the total value of the collateral is higher than the total value of the circulating stablecoin supply and is applied to protect the peg against the volatility of crypto tokens and related sudden price changes of the collateral. Any DeFi user can mint the new stablecoins, by providing sufficiently over-collateralized collateral. The process can also be seen as borrowing stablecoins with crypto tokens locked as collateral.

The borrowing process from the stablecoin protocol is initiated by depositing crypto tokens as collateral. The collateralized debt position (CDP) is created. In exchange for the deposit, the depositor receives stablecoins, which are minted by the stablecoin protocol. Once the borrower returns the stablecoins, the CDP is closed: the tokens used as collateral are returned to the borrower and the stablecoin protocol burns returned stablecoins.



\paragraph{Algorithmic stablecoins}
Algorithmic stablecoins maintain their price peg by managing (shrinking or expanding) the supply of the token. As they do not employ any collateral, algorithmic stablecoins are also referred to as non-collateralized stablecoins. There are two main types of algorithmic stablecoins: seigniorage and rebasing \cite{Mell2022UnderstandingConsiderations}. The rebasing coins, such as Ampleforth\cite{2023Ampleforth}, adjust the balance of coins in the users account, allowing the coin base to fluctuate. Consequently, some literature does not consider the rebasing coins to be stablecoins. Seigniorage stablecoin protocols, such as Terra\cite{2023Terra}, make arbitrary decisions on the minting and burning of the coins in the supply in order to maintain the peg. 

\paragraph{Risks}
\color{black}
\emph{De-peg risk} - the major risk related to the stablecoin - references to the circumstances in which the stablecoin loses the peg to the target value. It can be caused by the loss of the collateral underwriting the stablecoin in the aftermath of the hacking attack or insolvency of the custodian bank, in the case of the fiat-backed stablecoins. For instance, USDC - a stablecoin pegged to 1 USD, encountered 10\% de-peg over a couple of days due to the insolvency risk of Silicon Valley Bank, in which 10\% of USDC reserves were under custody \cite{2023CircleCoin}.
According to the \emph{stablecoin trilemma}, any stablecoin design poses limitations related to one of the following risks: \1 hazards of the operating entity, \2 exposure to the external market risk, or \3 limited coin supply. Due to their design, crypto-backed stablecoins cannot optimize the coin supply, whereas the fiat-back stablecoins suffer the hazards of the operating entity \cite{Kwon2021TheStablecoin}.

The major risk of DeFi lending, including borrowing from interest-rate or stablecoin protocols, is the \emph{liquidation risk}. Liquidation occurs when the value of the collateral drops below the minimum collateralization threshold. Borrowers are required to monitor the over-collateralization level of their CDP to avoid liquidations. The automatic detection of the CDP that needs to be liquidated and initiating the liquidations are not trivial on the blockchain. Smart contacts, by their design, need to be triggered externally and do not support automation.
During the market turmoil of high price volatility, the stablecoin protocol risks that the liquidation of the under-collateralized debt position is executed too late and below the debt value, leading to the protocol's losses and de-peg risk. There are various mechanisms applied by stablecoin protocol to cover the losses of the liquidations, \eg with the sale of governance tokens \cite{MakerDAOTeam2017TheSystem} or with other debt positions \cite{2023Liquity}.
Crypto-backed stablecoins might also be affected by \emph{oracle attacks}, in which the price feeds used by the protocol are manipulated. Most crypto-backed stablecoins are pegged to 1 USD \cite{2023CoinGecko}, and the exchange rate between 1 USD and the underlying crypto-currencies used as collateral is a critical component of their operations. The oracle price manipulations are further discussed in section \ref{sec:risk}.
\color{black}

\paragraph{CBDC}
Stablecoins are typically discussed in relation to the topic of \emph{Central Bank Digital Currencies} (CBDC), as a potential implementation of CBDCs. CBDC is a form of digital cash issued by the central bank. CBDC may be but does not have to be related to the BC or DLT, as it is the issuer - the central bank -  of digital cash that is a defining factor of CBDC. CBDCs are further classified as wholesales or retail\cite{Chaum2021HowCurrency}. The potential CBDC implementation may be decentralized or centralized, and may not relay on DLT\cite{Chaum2021HowCurrency}. 

\subsection{Liquidity Pool}

\subsubsection{Interest-rate Protocols}
Interest rate protocols, also referred to as lending protocols, facilitate borrowing and lending of digital assets \cite{Chiu2022OnLending, Saengchote2022DecentralizedCompound, Lehar2022SystemicMarkets}. Lenders - liquidity providers (LPs), provide capital to the protocol's pool - \emph{lending pool}, from which it is distributed to borrowers. The interest rate mechanism seeks for equilibrium between the demand and supply of the capital \cite{Gudgeon2020DeFiEfficiency}. DeFi lending protocols are thus not decentralized peer-to-peer lending strategies with direct relations between borrowers and lenders. Instead, they operates in a peer-to-pool model that avoids individual matching. Thus, the LP has a diversified risk across all borrowers of this pool and a utilization rate $U_t$ can be calculated at time $t$. The utilization rate is the ratio of the toal amount borrowed at time $t$ to the liquidity at time $t$, so $U_t := \frac{B_t}{L_t}$\cite{2020AaveV1.0}. Liquidity is provided to the lending pool managed by the smart contract of the protocol. Thus, the DeFi lending protocols are also referred to as \emph{protocols for loanable funds} (PFLs)\cite{Gudgeon2020DeFiEfficiency}. Unlike in traditional banking, DeFi participants are anonymous, preventing assessing the risk of borrowers \cite{Aramonte2022BISInformation}. To mitigate the default risk, DeFi interest-rate protocols are over-collateralized: a borrower is required to post collateral covering the value of the debt and the value of the posted collateral must exceed the value of the debt. If the value of the locked collateral falls below the \emph{liquidation threshold}, the collateral can be purchased for a discounted value and the debt position is closed \cite{Perez2020Liquidations:Knife-edge}. DeFi loans typically have unlimited maturities and the interest rate is variable, so it fluctuate from the opening of the loan position and is recalculated constantly. Interest is accrued per second and paid per block\cite{Gudgeon2020DeFiEfficiency}. Based on the interest rate model, the lending protocol can be classified into three major classes: linear, non-linear and kinked rates.
    
    \paragraph{Linear rates}
    In a linear model, interest is a linear function of the utilization rate. For the borrower, the variable interest rate $i_{b,t}$ results from: $$i_{b,t} = \alpha + \beta \cdot U_t,$$ where $\alpha$ is a constant and $\beta$ describes how the interest rate behaves for changes in the utility rate. The lender receives a lower saving interest rate $i_{s,t} = i_{b,t}\cdot U_t$ which also depends on $U_t$ \cite{Gudgeon2020DeFiEfficiency} For example, the first version of Compound used this model, but it has since been replaced by a kinked rate model \cite{RobertLeshner2019Compound:Protocol}. 
    
    \paragraph{Non-linear rates} In a non-linear model interest rates increase with an increasing rate as the utilization grows. The interest rate structure was used in the derivatives platforms dYdX (v1, v2) and DDEX\cite{2023DDEX}, where the borrowing interest rate is given by $$i_{b,t} = (\alpha\cdot U_t)+(\beta \cdot U_t^{n})+(\gamma\cdot U_t^{m}),$$ with $n,m \in \mathbb{R}$ The saving interest rate $i_{s,t} = (1-\lambda)\cdot i_{b,t}\cdot U_t$ includes a reserve factor $\lambda$, which puts aside a fraction of the earned interest rate for more illiquid periods \cite{Gudgeon2020DeFiEfficiency}. 
    
    \paragraph{Kinked rates} In the kinked model, interest rates increase more sharply above a certain (optimal) utilization rate $U_{opt}$, i.e. there is a kink in the interest rate slope. This is used in the current version of AAVE  \cite{2020AaveV1.0} and Compound \cite{RobertLeshner2019Compound:Protocol}. The borrower rate can be characterized by $$ i_{b,t}= \begin{cases} \alpha + \beta U_t &\text{, if } U\leq U_{opt} \\ \alpha + \beta U_{opt} + \gamma (U-U_{opt}) &\text{, if } U> U_{opt}\end{cases}, $$ where $\beta$ describes the interest rate slope up to the optimal utilization rate and $\gamma >> \beta$ the slope for higher utilization. 
\paragraph{Total value locked} (TVL) for the interest-rate protocols represents the value of tokens hold by the protocols, which is the difference between the supplied and borrowed funds, and the value of the collateral. The major interest-rate protocols, Aave\cite{2020AaveV1.0} and Compound\cite{RobertLeshner2019Compound:Protocol} with  USD 3.37bn and USD 1.64b TVL respectively\cite{2022DeFiLlama}, are examples of DeFi interest-rate protocol with kinked rates. 
\color{black}
The perpetual exchange dYdX\cite{Juliano2017DYdX:Derivatives} implemented non-linear rates to offer leverage to its users.

\paragraph{Risks}
Interest rate protocols require a borrower to post collateral with a value exceeding the value of the debt. This mechanism incentives a borrower to repay a loan and protect the capital of LPs, though it also exposes the borrower to the \emph{liquidation risk}.
In case the value of the collateral drops below the liquidation threshold, the borrower automatically defaults on its debts: the collateral is sold off with the discount, in the process referred to as \emph{liquidation}. Whereas the liquidation risk primarily affects the borrower, during the black swan event of extreme market volatility, the liquidation of the collateral might be executed below the value of the debt, resulting in losses for the LPs. To prevent this situation, the interest rate protocol sets various liquidation threshold above the debt value. Additionally, the liquidation fee is charged from the liquidated collateral \cite{Perez2020Liquidations:Knife-edge, Gudgeon2020DeFiEfficiency}. \emph{Zero liquidation loans} (ZLLs) are the novel approach to mitigate the DeFi liquidation risk by defining the loan tenor (duration) \cite{Sardon2021Zero-LiquidationLending}. The ZLLs protocols such as \cite{SaradonA.2022MYSOLoans, Ngo2021TimeswapProtocol, 2023VendorFinance} offer the borrower an option (but not obligation) to reclaim the collateral upon the repayment of the debt and the fees. If a borrower does not repay the debt before the ZLL expires, LPs gain the right to liquidate the collateral. 

\emph{Oracle attack} refers to manipulating the price feeds used by the DeFi protocols. Interest rate protocols depend on the exchange rates between tokens, and consequently, they are susceptible to price manipulation. For instance, when the exchange price between the token of collateral and the token of debt is manipulated, the protocol cannot calculate the correct collateral-debt ratio and can trigger liquidation. Oracle attacks are further discussed in section \ref{sec:risk}.

\paragraph{Flash loans}
Flash loans are a special class of DeFi loans that allow to borrow without collateral during a single transaction. The debt is protected by the atomicity enforced by smart contracts and blockchain: if the loan is not repaid with interest within the block, the whole transaction with a flash loan is reversed\cite{2020AaveV1.0}. 
Flash loans increase the efficiency of DeFi lending by allowing to purchase any digital assets that can be used as collateral, including NFT \cite{BakerLucas2022ParadigmsCredit}, which presents algorithm \ref{alg:NFTFlashLoan}. The platforms \cite{2023NFTFi, 2023BendDAO} accept certain NFTs as collateral.
\color{black}
\begin{algorithm}
\caption{Mortgaging NFT\cite{BakerLucas2022ParadigmsCredit}}\label{alg:NFTFlashLoan}
\begin{algorithmic}[1]
\STATE \textbf{Flash loan} $x_i$ of token $i$ to an address holding $y_i$ of token $i$
\STATE \textbf{Purchase} the NFT for $x_i+y_i$ of token $i$
\STATE \textbf{Deposit} the NFT as collateral, with borrowing power of $z_i > x_i$
\STATE \textbf{Borrow} $x_i$ of token $i$ to repay the flash loan
\end{algorithmic}
\end{algorithm}

\subsubsection{Decentralized Exchanges}

Decentralized Exchange (DEX) is a class of DeFi Protocols that facilitates the non-custodial swap of digital assets with on-chain settlement \cite{Werner2021SoK:DeFi} \cite{Lin2019DECONSTRUCTINGEXCHANGES} and is the largest in terms of TVL class of DeFi protocols\cite{2022DeFiCategories}. Depending on the price discovery mechanism, DEXs can be classified as \emph{Central Limit Order Book} (CLOB)-DEX, or \emph{Automated Market Maker} (AMM)-DEX, \cite{Jensen2021AnDeFi}. CLOB proved to be infeasible and expensive at scale on the blockchain\cite{Jensen2021AnDeFi}. 

AMM-DEX employs reserves of two or more tokens, called \emph{liquidity pools}. Liquidity pools provide counterparty tokens to swap against  - and the swap price is determined algorithmically by the pricing rules called \emph{Automated Market Maker} (AMM). AMM-DEX became dominant in DeFi \cite{Werner2021SoK:DeFi}. Depositors that seed liquidity pools with capital in return for swap fees, are referred to as \emph{liquidity providers} (LPs). The trading fees are distributed to LPs proportionally to the capital provided to the liquidity pool. The process of providing liquidity to the DEX, or other class of DeFi protocols, in return for profit, is similar to market making in traditional finance. Market makers in traditional finance provide both bid and ask orders from their own book and make a profit from the bid-ask spread. 

\emph{Arbitrageurs} buy and sell the same asset at various exchanges to profit from the price differences. Even though small liquidity pools may lead to significant fluctuations in the swap prices, arbitrageurs ensure that the swap price at the DEX is at parity with the market. The next section presents the most used AMM.

\paragraph{Automated Market Maker (AMM)}: 
An AMM is a mathematical function that algorithmically determines the swap price between tokens in the liquidity pool via a scoring rule. The most popular scoring rule is the logarithmic market scoring rule from Robin Hanson \cite{Hanson2002LogarithmicAggregation}. In recent years new AMMs emerged in the field of decentralized finance, the \emph{constant function market makers} (CFMM). A CFMM is a trading function $\varphi: \mathbb{R}^n_+ \rightarrow \mathbb{R} $, often called a conservation function or reserve curve, that maps the token reserves $R_i$ to an constant $k$, often referred to as invariant. 
Therefore, the conservation function is a relational function between an invariant $k$ and token reserves $R_i$. In addition, the conservation function is always concave, nonnegative, and nondecreasing\cite{Angeris2021AMaximization} and the spot price is given by the slope of the curve at the current reserves level $ p_{R_i,R_j} = {\frac{\partial \varphi}{\partial R_i}}/{{\frac{\partial \varphi}{\partial R_j}}}$. If the liquidity pool contains only two tokens, the two reserves are typically referred to as $x$ and $y$. Various AMM formulas are currently in use by DEXs. The most popular AMMs fall into the category of \emph{constant function market making} (CFMM),  and include the following \emph{reserve curves} \cite{Xu2021SoK:Protocols, Neuder2021StrategicV3}:

 \paragraph{Constant Sum Market Maker}
    \[ x + y = k \] The constant Sum Market Maker (CSMM) is the simplest form of an AMM. It sums over all reserves and the sum is defined as the invariant $k$, resulting in a swap exchange rate of 1. This means that a small deviation from the price will lead to arbitrage and one token reserve will be completely drained. Therefore, this AMM is unsuitable as a standalone solution for DEXs.
    
    \paragraph{Constant Product Market Maker}
    \[ x \cdot y = k \] The Constant Product Market Maker (CPMM) is the most known DeFi AMM and became popular through its use in Uniswap v2. The function defines the interaction of two token reserves and the invariant by a simple product of the two reserves. In a swap operation, this always remains the constant $k$,  so $\Delta x$ of $x$ can be swapped for $\Delta y$ of $y$, resulting in $(x+\Delta_x)(y-\Delta_y)=k$ and a spot exchange rate of $p_{x,y} = \frac{x}{y}$.

    \paragraph{Constant Mean Market Maker}
    \[ \prod\limits_{k=1}^n R_i^{w_i}\] The Constant Mean Market Maker (CMMM) is the extension of the CPMM. This AMM includes more than 2 token reserves $R_i$ and weights them with $w_k$ and was first introduced by Balancer in 2019\cite{MartinelliF2019BalancerSensor.}. It results in an spot exchange rate of $p_{i,j} = \frac{R_i\cdot w_j}{R_j\cdot w_i}$.

\color{black}

    \paragraph{Stableswap Invariant}
    The Stableswap Invariant \cite{Egorov2019StableSwapLiquidity}, often also referred to as Curve v1 AMM, mixes the behavior of the CPMM and CSMM and is mostly meant for Stablecoins. The invariant is defined as: 
     \begin{equation}\label{eq:curvev1}
        B\cdot k^{n-1}\cdot \sum\limits_{i=1}^n R_i +\prod\limits_{i=1}^n R_i = B\cdot k^n + (\frac{k}{n})^n
    \end{equation} with a dynamic leverage factor $B= \frac{A\cdot\prod_{i=1}^n R_i}{k^n}\cdot n^n$ which is equal to a constant $A$ as long as the price is equal to the stable price and goes to zero otherwise. Substituting $B$ in equation (\ref{eq:curvev1}) simplifies to:
        
  \[ A\cdot n^n\cdot \sum\limits_{i=1}^n R_i +k = A\cdot k\cdot n^n + \frac{k^{n+1}}{n^n\prod_{i=1}^n R_i}\]

   The conservation function includes a leverage/amplification coefficient $A \in \mathbb{R}_+$ which describes how strongly the function follows a CPMM or CSMM. For $A \rightarrow 0$ the function behaves like the CPMM, for $A \rightarrow \infty$ like the CSMM. The spot exchange price is given by $$p_{i,j} = \frac{R_i\cdot (A\cdot R_j \cdot \prod_{l=1}^n R_l + \frac{k^{n+1}}{n^n})}{R_j\cdot (A \cdot R_i \cdot \prod_{l=1}^n R_l + \frac{k^{n+1}}{n^n})}$$
    \paragraph{Cryptoswap Invariant}
The Cryptoswap invariant \cite{Egorov2021CurvePeg}, also known as Curve v2, is the further development of the stableswap AMM, which can be used for non-stablecoins. The Cryptoswap invariant requires adjusting the leverage factor $B$ in equation (\ref{eq:curvev1}) to: 
$$B = A\cdot \underset{\text{ =: }B_0}{\underbrace{\frac{A\cdot\prod_{i=1}^n R_i}{k^n}\cdot n^n}} \cdot \frac{\gamma^2 }{(\gamma+1- B_0)^2},$$ where $\gamma$ describes the distance between two CPMMs\footnote{The two CPMMs are given by $\prod R_i = (\frac{k}{n})^n$ and $\prod R_i = \gamma(\frac{k}{n})^n$} in which the conservation function is located. Another innovation is, that  curve v2 concentrates the liquidity around an internal price in such a way that the repegging costs and fee revenues are optimised. For this purpose, an exponential moving average is used as a price oracle $p^*$ and the value of the portfolio is defined as $X_{cp} = (\prod_{i=1}^n \frac{k}{n\cdot R_i})^{1/n}$. Then the change in the value of the pool $X_{cp}$ is compared to the profits of the pool and as soon as the losses exceed half of the profits, the peg is adjusted towards $p^*$.\\

\begin{figure*}[htp]
\begin{minipage}[c]{0.45\linewidth}
\centerline{\includegraphics[width=0.95\linewidth]{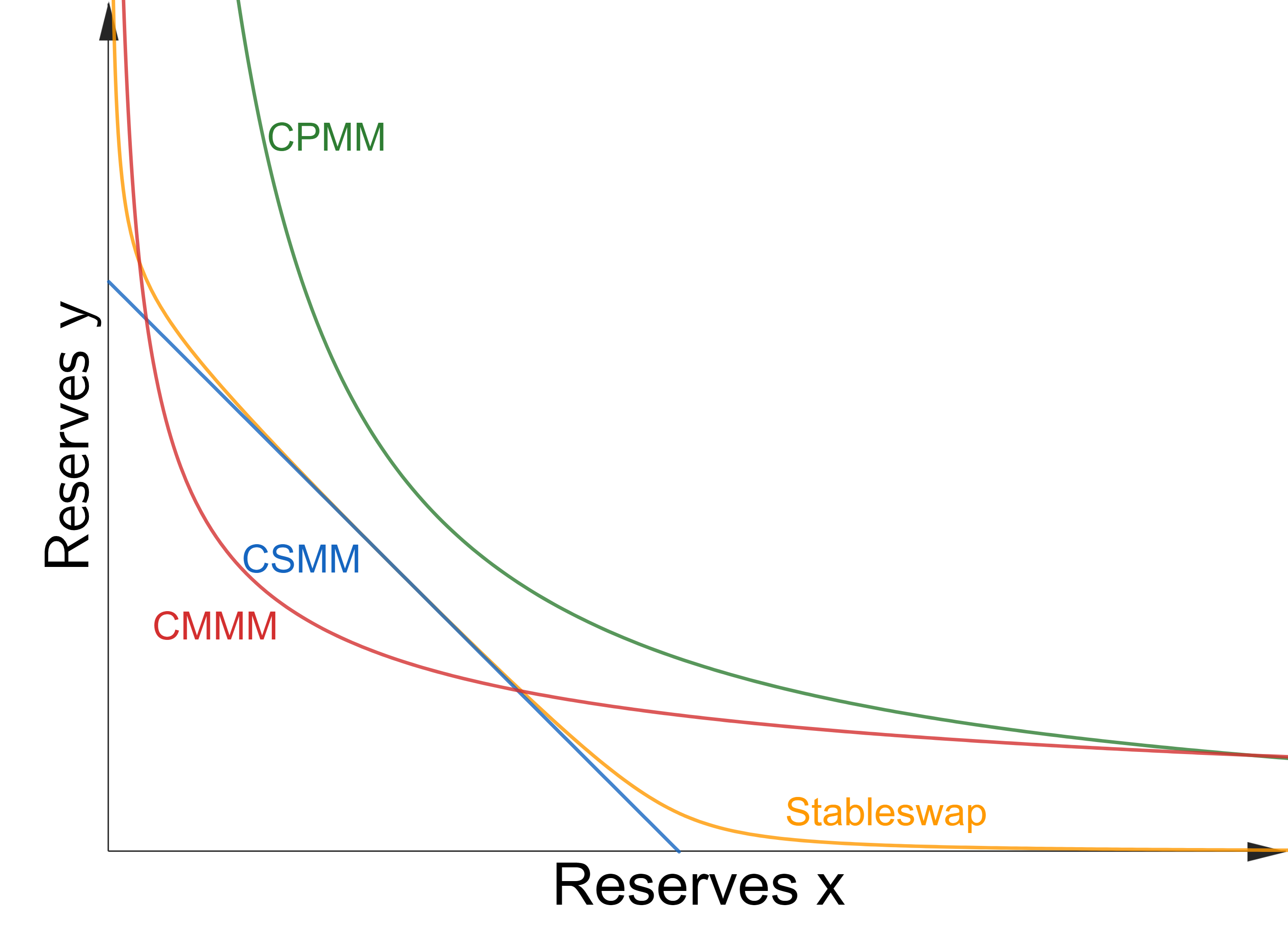}}
\caption{Graphical representation of the AMM conservation functions for two tokens with CMMM weights $w_x =1, w_y=2$ and a stableswap leverage coefficient of $A=10$}
\label{fig:consfunc}
\end{minipage}\hfill
\begin{minipage}[c]{0.45\linewidth}
\centerline{\includegraphics[width=0.95\linewidth]{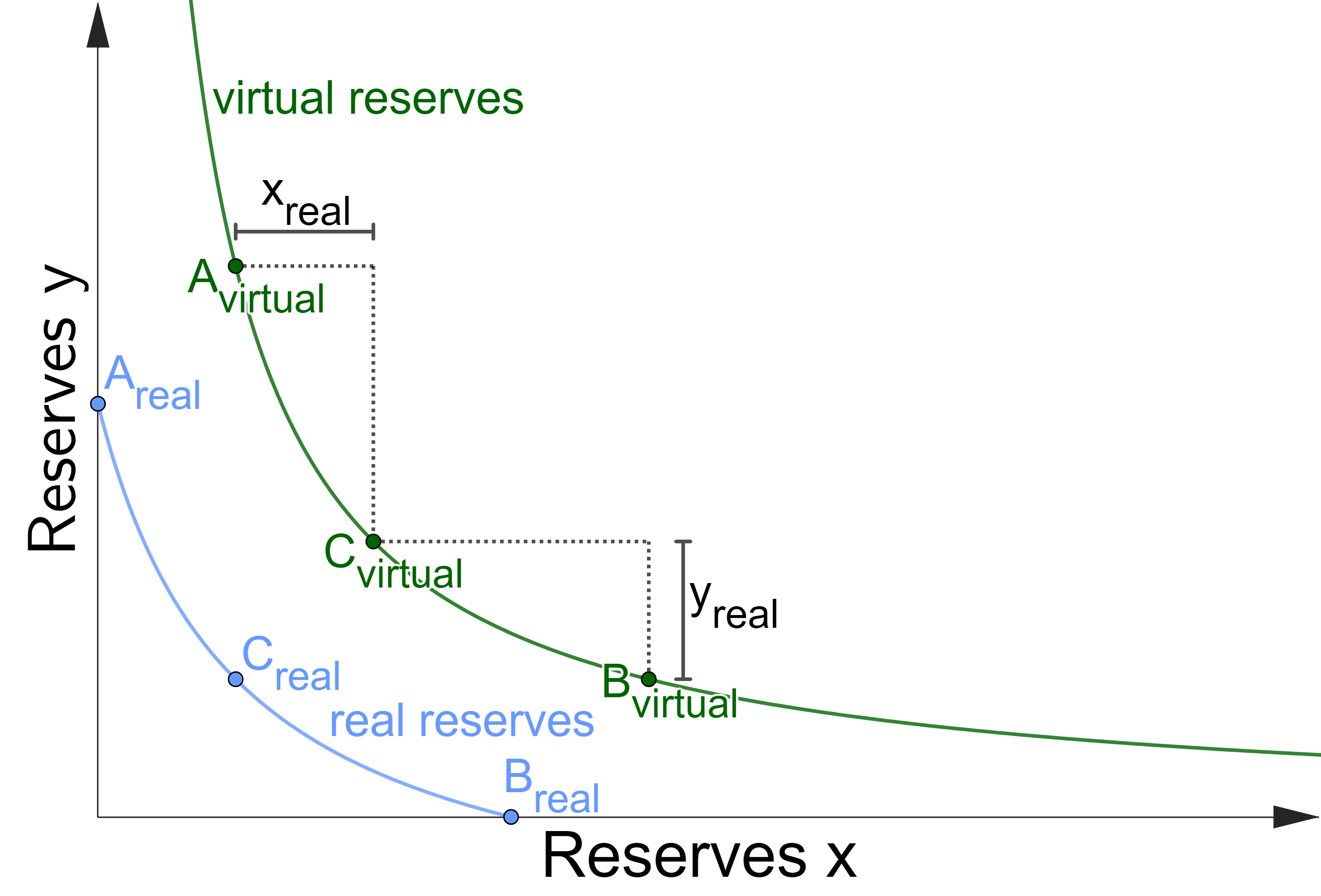}}
\caption{Concentrated liquidity and virtual reserves, based on \cite{Adams2021UniswapCore}}
\label{fig:uniswap3}
\end{minipage}
\end{figure*}

Figure \ref{fig:consfunc} gives a graphical representation for each of these conservation functions except the Cryptoswap, which has the same form as the Stableswap. 

\paragraph{Concentrated Liquidity}
Concentrated Liquidity AMM (CLAMM) is an enhancement to AMM that increases capital efficiency in liquidity pools\cite{2021ProgrammingUniV3}. It was first proposed by UniSwap v3\cite{Adams2021UniswapCore} and allows LPs to define the price range, for which liquidity is provided. LPs profit from swap fees when the price remains in the defined interval\cite{2021ProgrammingUniV3}. Consequently, there is a trade-off between larger (less risky, less efficient) intervals and (more risky, efficient) smaller intervals. Consider a liquidity range interval $[p_a,p_b]$ around the current spot price $p_c$, then $x_{real}$ and $y_{real}$ denotes the position real reserves. As soon as the price moves outside the range, a reserve of the liquidity provider is completely used up. Consequently, CLAMM requires active management and accepting higher financial risk, compared to AMM \cite{Heimbach2022RisksProviders}. The liquidity provided is measured by the invariant L, which equals $\sqrt{k}$ in the CPMM setting. The real reserves of a position are described by a translation of the standard CPMM formula: $$ (x+\frac{L}{\sqrt{p_b}} )(y+L\cdot \sqrt{p_a}) = L^2$$ Therefore, the spot exchange rate is identical to the CPMM case ($p_{x,y} = \frac{x}{y}$). Figure \ref{fig:uniswap3} shows the real and virtual reserves and the shift in the curve. 


\paragraph{Risks}
The major risks for the participants of AMM-based DEX include 1) slippage cost for traders, 2) impermanent loss for LPs and 3) transaction-ordering attacks. Those risks come additionally to the explicit costs of DEX - swap fees and gas fees. Transaction reordering, or MEV-attack, is further discussed in this work, as it poses a risk for many classes of DeFi protocol.

\emph{Slippage} is the difference between the spot price, presented by the DEX, and the actually realized swap price. The difference is caused by AMM and the reserve curves. The realized price diverges from the spot price for larger swapped amounts and the divergence is further amplified by smaller liquidity pools. Swap transactions at AMM DEX are set with the \emph{slippage tolerance}. Slippage tolerance restricts losses from the MEV attacks.

\emph{Impermanent loss} is the difference between holding tokens in the wallet and allocating them to the liquidity pool. Every swap transaction alters the liquidity pool's composition, increasing the value-losing token's reserve.
Impermanent loss can be eliminated by providing liquidity for stablecoin pairs \cite{Lipton2021AutomatedCurrencies} or mitigated by including stablecoin in the token pair \cite{Heimbach2021BehaviorExchanges}. The impact of impermanent loss has been widely studied across various AMMs\cite{Labadie2022ImpermanentFormula, Tiruviluamala2022AMakers}.

\subsubsection{Perpetuals}
Perpetual contracts or swaps in DeFi are derivatives that allows to speculate on the price of an underlying asset, such as cryptocurrencies, without actually owning the asset. Perpetual contracts resemble traditional futures or options, however, they do not have a fixed expiration date. This allows to hold their positions indefinitely, as long as the required collateral limits can be maintained. They were first introduced by the centralized exchange BitMEX in 2016 \cite{Alexander2019BitMEXEffectiveness}, requiring counter-parties and thus a matching mechanism (similar to CLOB).

While automatic liquidation of collateral is an essential mechanism for any perpetual, another fundamental DeFi mechanism is necessary to remove the need for counter-parties: virtual AMMs (vAMM). A vAMM can use the same mathematical principles as an AMM for price determination, which is required for automatic liquidation of collateral once the price hits a certain limit, \emph{e.g.}, if the price shifts to your disadvantage, you may require to provide more collateral. In case your collateral limit is reached (also depends on your leverage factor), your collateral will be automatically liquidated.



\subsection{Aggregators}

\subsubsection{DEX Aggregators}
The rise of DEXs on the same blockchain provides more competitive options for DeFi users, but also leads to fragmented liquidity among various AMM pools and requires arbitrageurs for price synchronization. DEX arbitrage and trade routing can be done efficiently on-chain within smart contracts\cite{Zhou2021A2MM:Exchanges}. \emph{DEX aggregators} provide the optimal routing of trade orders across multiple DEXs \cite{Angeris2022OptimalMakers}. Around 21.5\% of DEX volume is initiated by aggregators\cite{2023DuneMetrics}. 

\subsubsection{Yield Farming Aggregators}
Yield farming is the process of earning income by providing tokens to DeFi protocols for market-making purposes\cite{Cousaert2022SoK:DeFi} while token providers are referred to as "yield farmers". The yield farming protocols automate the yield farming process and can be perceived as algorithmic asset managers that execute the pre-programmed strategy \cite{Xu2022AProtocols}. Technically, yield farming protocols are aggregators that automatically deploy tokens to other DeFi protocols \cite{Xu2022ReapProtocols} and are consequently also referred to as "yield aggregators". 
The income from yield farming activities might be amplified by the participation rewards from DeFi protocols, to which tokens are provided. Such participation rewards are typically offered as governance tokens. Yield farming protocols may charge management and performance fees.
\paragraph{Simple lending} In this strategy, the tokens are provided to the interest rate protocols (\eg Compound, Aave), which lend them to the borrower. The token provider (supplier) accrues the supply interest. Given the over-collateralization required by the interest rate protocols from the borrowers, the strategy is low-risk and losses occur only under extreme market conditions.
Based on the source of yield, the protocol strategies can be divided into the three groups\cite{Cousaert2022SoK:DeFi}: \1 simple lending, \2 leverage borrow, \3 liquidity provision.

\begin{algorithm}
\caption{Simple Lending Yield Farming Strategy\cite{Cousaert2022SoK:DeFi}}\label{alg:alg1}
\begin{algorithmic}[1]
\STATE \textbf{Deposit} token $i$ in a lending protocol
\STATE \textbf{Accrue} interest of token $i$ and collect potential other rewards, e.g. native tokens $z$ over time
\STATE \textbf{Withdraw} deposits with accrued supply interest and other rewards
\end{algorithmic}
\end{algorithm}

\paragraph{Leverage borrow} This strategy aims to maximize the yield from lending by leveraging spirals: tokens deposited in the interest rates protocols are used as collateral to borrow new tokens. The new tokens are deposited in the interest rate protocol and used again as collateral to borrow. The process is repeated multiple times. A high degree of leverage can amplify both profits and losses.

\begin{algorithm} 
\caption{Leverage Borrow Yield Farming Strategy\cite{Cousaert2022SoK:DeFi}}\label{alg:alg1}
\begin{algorithmic}[1]
\STATE \textbf{Deposit} token $i$ in a lending protocol
\STATE \textbf{Borrow} token $j$ with the deposits as collateral
\STATE \textbf{Deposit} borrowed token $j$ as collateral
\STATE \textbf{Repeat} steps 2–3 multiple times
\STATE \textbf{Accrue} interest and collect potential other rewards, e.g. native tokens $z$ over time.
\STATE \textbf{Swap} the native tokens into the assets borrowed
\STATE \textbf{Repay} loans with accrued borrow interest
\STATE \textbf{Withdraw} deposits with accrued supply interest and other rewards

\end{algorithmic}
\end{algorithm}

\paragraph{Liquidity provision} With this strategy, tokens are provided to liquidity pools (LPs) of AMM DEXs to earn income in terms of transaction fees charged by DEXs. 

\begin{algorithm}
\caption{Liquidity Provisions Yield Farming Strategy\cite{Cousaert2022SoK:DeFi}}\label{alg:alg1}
\begin{algorithmic}[1]
\STATE \textbf{Provide} token $i,j$ as liquidity in an AMM pool
\STATE \textbf{Collect} transaction fees in token $i,j$ and potential other rewards, e.g. native tokens over time
\STATE \textbf{Withdraw} liquidity, accrued fee participation and other rewards
\end{algorithmic}
\end{algorithm}


\section{DeFi Risk Classification}
\label{sec:risk}

With the rapid increase of TVL, attacks on DeFi protocols have gradually emerged \cite{Li2022SecurityAdvances, Zhou2022SoK:Attacks, Chen2019ADefenses}. In \emph{DeFi exploits}, attackers gain profits from the vulnerability in the technical or economic security of the DeFi protocols \cite{Werner2021SoK:DeFi}. Table \ref{tabl:DeFiRisk} summarizes DeFi risks and their impact on DeFi stakeholders - service customers, liquidity providers (LPs), arbitrageurs, or governance users. This section discusses these risks and attacks in detail. As yield farming protocols are aggregators that provide liquidity to other DeFi protocols, this category possesses risks of LPs.

\begin{table*}[ht]
\caption{DeFi Security Risk Overview. \\ C: Service Customers, LP: Liquidity Provider, A: Arbitrageur, G: Governance User  \\ \emptycirc = Not Vulnerable, \fullcirc = Vulnerable, \halfcirc = Depends}
\begin{center}

\begin{tabular}{l|lll|lll|lll|ll|ll|ll|l}
                               & \multicolumn{3}{l|}{\textbf{CLOB}}    & \multicolumn{3}{l|}{\textbf{AMM}}                                                  & \multicolumn{3}{l|}{\textbf{Interenst-rate}} & \multicolumn{2}{l|}{\textbf{Stablecoins}}     & \multicolumn{2}{l|}{\textbf{Bridges}}          & \multicolumn{2}{l|}{\textbf{Liquid}}  &            \\
                               & \multicolumn{3}{l|}{\textbf{DEX}}     & \multicolumn{3}{l|}{\textbf{DEX}}                                                  & \multicolumn{3}{l|}{\textbf{Protocols}}      & \multicolumn{2}{l|}{\textbf{}} & \multicolumn{2}{l|}{\textbf{}} & \multicolumn{2}{l|}{\textbf{Staking}} &            \\ \hline
                               & \textbf{C} & \textbf{LP} & \textbf{A} & \textbf{C} & \textbf{\begin{tabular}[c]{@{}l@{}}LP\\ YF\end{tabular}} & \textbf{A} & \textbf{C}    & \textbf{LP}   & \textbf{A}   & \textbf{C}            & \textbf{A}            & \textbf{C}             & \textbf{A}            & \textbf{C}         & \textbf{A}       & \textbf{G} \\ \hline
\textbf{DeFi Risks}            &            &             &            &            &                                                          &            &               &               &              &                       &                       &                        &                       &                    &                  &            \\
Rug Pull Attack                &\fullcirc   &\fullcirc    &\fullcirc   &\fullcirc   &\fullcirc                                                 &\fullcirc   &\emptycirc     &\emptycirc     &\emptycirc    &\emptycirc             &\emptycirc             &\emptycirc              &\emptycirc             &\emptycirc          &\emptycirc        &\emptycirc  \\
Slippage                       &\emptycirc  &\emptycirc   &\emptycirc  &\fullcirc   &\emptycirc                                                &\fullcirc   &\emptycirc     &\emptycirc     &\emptycirc    &\emptycirc             &\emptycirc             &\emptycirc              &\emptycirc             &\emptycirc          &\emptycirc        &\emptycirc  \\
Impermanent Loss               &\emptycirc  &\emptycirc   &\emptycirc  &\emptycirc  &\halfcirc                                                 &\emptycirc  &\emptycirc     &\emptycirc     &\emptycirc    &\emptycirc             &\emptycirc             &\emptycirc              &\emptycirc             &\emptycirc          &\emptycirc        &\emptycirc  \\
Liquidation                   &\emptycirc  &\emptycirc   &\emptycirc  &\emptycirc  &\emptycirc                                                &\emptycirc  &\fullcirc      &\halfcirc     &\fullcirc     &\fullcirc              &\emptycirc             &\emptycirc              &\emptycirc             &\emptycirc          &\emptycirc        &\emptycirc  \\
De-peg                         &\emptycirc  &\halfcirc    &\emptycirc  &\emptycirc  &\halfcirc                                                 &\emptycirc  &\emptycirc     &\halfcirc      &\emptycirc    &\fullcirc              &\emptycirc             &\fullcirc               &\emptycirc             &\fullcirc           &\emptycirc        &\fullcirc   \\ \hline
\textbf{Infrastructure Attack} &            &             &            &            &                                                          &            &               &               &              &                       &                       &                        &                       &                    &                  &            \\
MEV Attack                     &\fullcirc   &\emptycirc    &\fullcirc   &\fullcirc   &\halfcirc                                                 &\fullcirc   &\emptycirc     &\emptycirc     &\emptycirc    &\emptycirc             &\emptycirc             &\emptycirc              &\emptycirc             &\emptycirc          &\emptycirc        &\emptycirc  \\
Blockchain Attack              &\halfcirc   &\fullcirc    &\halfcirc   &\halfcirc   &\fullcirc                                                 &\halfcirc   &\fullcirc      &\fullcirc      &\halfcirc     &\fullcirc              &\halfcirc              &\halfcirc               &\halfcirc              &\fullcirc           &\halfcirc         &\fullcirc   \\ \hline
\textbf{Middle Layer Attack}   &            &             &            &            &                                                          &            &               &               &              &                       &                       &                        &                       &                    &                  &            \\
Smart Contract Attack          &\fullcirc   &\fullcirc    &\fullcirc   &\fullcirc   &\fullcirc                                                 &\fullcirc   &\fullcirc      &\fullcirc      &\fullcirc     &\fullcirc              &\fullcirc              &\fullcirc               &\fullcirc              &\fullcirc           &\fullcirc         &\fullcirc   \\ \hline
\textbf{Application Attack}    &            &             &            &            &                                                          &            &               &               &              &                       &                       &                        &                       &                    &                  &            \\
Oracle-Attack                  &\emptycirc   &\emptycirc    &\emptycirc   &\emptycirc   &\emptycirc                                                 &\emptycirc   &\fullcirc      &\fullcirc      &\fullcirc     &\fullcirc              &\fullcirc              &\emptycirc               &\emptycirc              &\emptycirc           &\emptycirc         &\fullcirc  
\end{tabular}

\label{tabl:DeFiRisk}
\end{center}
\end{table*}


The blockchain trilemma states that every BC can prioritize only two from three factors: decentralization, security, and scalability\cite{Conti2019BlockchainMessages}. The limited scalability of the BC results in network congestion and high transaction fees. A lack of decentralization allows for transaction censorship. The security attacks target vulnerabilities in the technical (\eg smart contract, transaction re-ordering) or economical design (\eg oracle price manipulation) of DeFi protocols and BCs. 

\begin{itemize}

\item \textbf{Limited Scalability} 
 Scalability is the central issue for BC networks\cite{Gangwal2022AProtocols}. The limited scalability leads to network congestion and spikes in transaction costs. During the network congestion, transactions of arbitrageurs, and other DeFi stakeholders, may not be executed, leading to market inefficiency and information delays\cite{Jensen2021AnDeFi}. Layer-2 solutions aim to scale the underlying main chain at the cost of various security assumptions\cite{Gangwal2022AProtocols}.

\item \textbf{Centralization Risk} 
Centralization might affect both the BCs and the DeFi protocols. DeFi Protocols increasingly rely on governance tokens to decentralize the decision-making process\cite{Jensen2021HOWFINANCE}. Nonetheless, the distribution of governance tokens may remain concentrated within the small groups of colluding stakeholders, as the distribution of voting powers is a gradual process\cite{Tsoukalas2020Token-weightedCrowdsourcing}.

\item \textbf{Security Threats}
DeFi Protocols operate on BC as smart contracts and are prone to security attacks on various layers: BC infrastructure, smart contract, and application attacks\cite{Xu2021SoK:Protocols}. 

\end{itemize}

\subsection{Infrastructure-layer Attacks}
The infrastructure-layer attacks include vulnerabilities in the BC design. They might directly target the BC, its nodes, and consensus mechanisms on which DeFi protocols operate (BC Attack) or exploit design faults in processing the transaction (MEV Attack).
\subsubsection{Blockchain Attacks}
The 51\% attack allows tampering with the ledger data by controlling over half of the network hash rate. Other examples include block timestamp manipulation.  DeFi service customers and arbitrageurs are partially affected during the BC attacks, as they are only restricted from using the DeFi Protocols. LPs lose all tokens locked on the victim BC. Similarly, issuers of stablecoins and liquids staking tokens lose the collateral. The governance users are fully affected, as their DeFi protocol is based on the victim BC. The BC attacks might be mitigated by using DeFi protocols only from the established BCs with vast networks of nodes and staked assets.
\subsubsection{Collapse of Terra Blockchain}
The Terra blockchain belonged to the ten largest blockchains with a market capitalization of over \$30bn. Its collapse was caused by the vulnerability in the economical design of its algorithmic stablecoin UST, \cite{Davernas2022Can}, and led to the fall of the DeFi protocols operating on Terra. In infrastructure layer attacks, such as the attack on Terra, all DeFi stakeholders - service customers, liquidity providers (LPs), arbitrageurs, and governance users - are fully exposed, as the attack ultimately leads to the loss of value of the victim blockchain with all deployed smart contracts and tokens.
\subsubsection{MEV Attacks}
\emph{MEV attacks} (Miner Extractable Value\cite{Daian2019FlashExchanges} attacks) refers to miners or validators benefiting from their access to information about an upcoming transaction. The traditional financial market has been encountering similar cases since the 1970s\cite{Eskandari2019SoK:Blockchain}. The blockchain, by design, allows validators to decide on the order of transactions in the block.  In MEV attacks, transactions are re-ordered for the benefit of validators at the expense of the initiator of the victim transaction. Flash loans additionally increase the arbitrage revenue of the attackers \cite{Qin2020AttackingProfit}. Some liquid staking protocols incorporate revenue from MEV attacks in the rewards tokens\cite{2023RocketPool}. Depending on the re-ordering strategy, the MEV attacks can be classified as follows:
\color{black}

\begin{itemize}
    \item \emph{Front-running attack} - the validator inserts the transaction before the victim transaction; if the victim transaction does not settle, the attack is fatal. This form of MEV attack targets, for instance, the arbitrage transaction. The validator makes a risk-free profit by copying and front-running the arbitrage transaction. The original transaction is not executed, as the arbitrage conditions no longer exist \cite{Yang2022SoK:Practice}
    \item \emph{Back-running attack} - the validator's transaction is executed after the target transaction. For instance, in MEV arbitrage, the validator executes the transaction that equals the prices between CEX and AMM-DEX \cite{Yang2022SoK:Practice}.
    \item \emph{Sandwich attack} combines front- and back-running, as depicted in algorithms \ref{alg:sandtwichAttack}, \ref{alg:sandtwichLPAttack}.
    \item \emph{Just In Time Liquidity} (JIT) attack occurs only at CLMM, such as Uniswap (v3), and refers to providing liquidity only for the price range of the target transaction. Once the transaction is executed, the liquidity is withdrawn from the pool, as presented in algorithm \ref{alg:JIT}.
\end{itemize}
\color{black}

\begin{algorithm}
\caption{Sandwich Price Attack\cite{Xu2021SoK:Protocols}}\label{alg:sandtwichAttack}
\begin{algorithmic}[1]

\STATE $User_A$ sets an order to buy token $i$ in exchange of token $j$ with current spot price $p_{i,j}$ on an AMM with gas fee $g_1$.
\STATE $User_B$ observes the mempool and sees the transaction.
\STATE $User_B$ front-runs the transaction by buying $x_B$ of token $i$ with a higher gas fee $g_2 > g_1$ on the same AMM.
\STATE $User_B$ and  $User_A$’s  transactions  are  executed  sequentially at respective average price of $p_{i,j}^B$ and $P_{i,j}^A$, pushing XYZ’s spot price up to $p_{i,j}^*$, where $p^*> p^A> p^B> p$ due to slippage.
\STATE $User_B$ back-runs by selling $x_B$ of token $i$ at an average price of $p^{B*}$, with $p^*> p^{B*}> p^B$ due to slippage

\end{algorithmic}
\end{algorithm}

DeFi protocols cannot prevent MEV attacks, as they are initiated in the blockchain infrastructure layer. Multiple approaches to mitigate transaction-ordering attacks have emerged, but all provide unsatisfactory results\cite{Heimbach2022SoK:Finance}. Considering swap fees, slippage tolerance, and gas fees, MEV attacks are only profitable if the size of the victim transaction exceeds a certain threshold\cite{Xu2021SoK:Protocols}. Privacy-preserving blockchains with TEE\cite{Zyskind2015DecentralizingData}, \eg Integritee\cite{2022Integritee} or Secret Network\cite{2022SecretNetwork}, prevent the transaction reordering from the infrastructure layer.

\begin{algorithm}
\caption{Sandwich LP Attack\cite{Xu2021SoK:Protocols}}\label{alg:sandtwichLPAttack}
\begin{algorithmic}[1]

\STATE $User_A$ places a transaction order to buy $x_A$ of token $i$ in exchange of token $j$ with gas  fee $g_1$ on an AMM with reserves $R_i,R_j$.
\STATE $LP_B$ observes the mempool and sees the transaction.
\STATE $LP_B$ front-runs by withdrawing liquidity from reserves $R_i$ and $R_j$ with a higher gas fee $g_2> g_1$.
\STATE $LP_B$ and $User_A$’s transactions are executed sequentially, resulting in a new composition of the pool. 
\STATE $LP_B$ back-runs by re-providing the liquidity for $R_i$ and $R_j$. Due to the transaction, the LP needs to provide less of token $j$ for the same share in the liquidity pool. 
\STATE $LP_B$  sells token $j$ for some token $i$ as profit. 

\end{algorithmic}
\end{algorithm}

\color{black}
MEV Attacks affect mostly traders (service customers) and arbitrageurs at CLOB and AMM DEXs. Traders pay the higher swap costs and arbitrageurs cannot execute their transaction. In JIT attacks, the traders profit from the attack, paying lower slippage costs. LPs are the victims of the attack, as their rewards from the swap fees are shared with the MEV attacker. Less than 1\% of liquidity at Uniswap (v3) is provided from JIT \cite{WanX.2022JustProtocol, XiongDemystifyingV3}.

\begin{algorithm}
\caption{Just in Time (JIT) Liquidity Attack\cite{WanX.2022JustProtocol, XiongDemystifyingV3}}\label{alg:JIT}
\begin{algorithmic}[1]

\STATE $User_A$ sets an order to swap token $i$ in exchange of token $j$ with current spot price $p_{i,j}$ on an AMM with gas fee $g_1$.
\STATE $User_B$ observes the mempool and sees the transaction.
\STATE $User_B$ front-runs the target transaction by providing liquidity to the pool through which the target transaction will be executed. The price range is set to the minimum range, in which the target transaction is executed.
\STATE $User_B$ executed the target transaction.
\STATE $User_B$ withdraws liquidity from the pool.

\end{algorithmic}
\end{algorithm}
\color{black}
\subsection{Smart-contract Layer-attacks}
Smart contract attacks exploit vulnerabilities in DeFi protocols caused by coding mistakes, unsafe calls to untrusted contracts, and access control mistakes. The audit of the DeFi protocol leads to a decrease in the probability of exploitation by a factor of four.\cite{Zhou2022SoK:Attacks}. \emph{Bug bounties} - reward programs for identifying software vulnerabilities - are a popular way to help to prevent smart contract exploits\cite{Breidenbach2018EnterContracts}. Smart-contract Layer attacks affect all DeFi stakeholders.


\color{black}
\subsection{Application Layer-attacks}
Application layer attacks seek to exploit vulnerabilities in the design of the DeFi protocols. One of the most notable examples of application layer attacks are oracle price manipulations.

\subsubsection{Oracle price manipulation}
Many DeFi protocols, \eg AMM-DEX or LSPs, do not require any market information to operate. Nevertheless, for some DeFi protocols, \eg interest-rate protocols and crypto-backed stablecoins, information from the real world are critical for the functionality. \emph{Oracles} are mechanism that retrieve the off-chain data, such as exchange prices between crypto-currencies and fiat currencies, which can be later utilized by smart contracts\cite{Liu2020AOracles}. There are various types of oracles. A centralized oracle relies on a trusted third party as a data provider, undermining the decentralization and trustlessness principles of DeFi. A decentralized oracle seeks the information on-chain, \eg from other DeFi protocols. For instance, a decentralized oracle may derive the swap rates between tokens from AMM-based DEX. This approach, however, is prone to the manipulations. The attack in which the prices used by the DeFi protocol are manipulated, is called an oracle attack. Algorithm \ref{alg:oracleattack} presents an oracle attack, in which AMM-based DEX is the source of the attack and the interest-rate protocols is the target of the attack. Oracle attacks affects all stakeholders of the targeted DeFi protocol.
\color{black}

\begin{algorithm}
\caption{Flash-loan-funded price oracle attack\cite{Xu2021SoK:Protocols}}\label{alg:oracleattack}
\begin{algorithmic}[1]

\STATE Take a flash loan to borrow $x_i$ of token $i$ from a lending platform, whose value is equivalent to $x_j$ of token $j$ at market price.
\STATE Swap $x_i$ of token $i$ for $x_j-\Delta_j$ of token $j$ on an AMM, pushing the spot price $p_{j,i}$ down. 
\STATE Borrow $x_i+ \Delta_i$ of token $i$ with $x_j - \Delta_j$ of token $j$ as collateral on a lending platform that uses the AMM as their sole price oracle. 
\STATE Repay the flash loan with $x_i$ of token $i$ and gain a profit of $\Delta_i$ of token $i$.

\end{algorithmic}
\end{algorithm}

\subsection{DeFi Risks}
DeFi stakeholders \ref{tab:DeFiAgents} are exposed to various DeFi risks, \eg liquidations, de-peg, impermanent loss, and slippage risk \ref{DeFiProtocolFundamentals} that are specific to the underlying DeFi algorithm. The risk impact differs between the stakeholders and affects in various levels service customers, LPs, arbitragers and governance users. Flash loans further amplify the losses of the victim and increase the profits of the attackers. 

\subsubsection{Rug  pull} A rug pull is a malicious activity in which developers abandon the project and use the capital raised from the investors as a profit \cite{Mazorra2022DoDetection}. It affects permissionless DEXs, which do not verify the listed tokens. At the interest rate protocols, crypto-backed stablecoins require the DAO vote to register new tokens as collateral, they are resistant to rug pull attacks. Bridges and liquid staking protocols depend only on the native tokens, which can not be rug pulled.

\begin{algorithm}
\caption{Rug Pull Attack\cite{Xu2021SoK:Protocols}}\label{alg:alg1}
\begin{algorithmic}[1]
\STATE \textbf{Mint} a new token $i$.
\STATE \textbf{Create} a  liquidity  pool  with $x_i$ of token $i$ and $x_j$ of a valuable token $j$ on an AMM.
\STATE \textbf{Attract} unwitting  traders  to  buy token $i$ with token $j$ from the pool, effectively changing the composition of the pool.
\STATE \textbf{Withdraw} liquidity from the pool, and obtain  $x_i - \Delta_i$ of token $i$ and $x_{j} + \Delta_j$ of token $j$ to take a profit of token $j$ of the valuable token.

\end{algorithmic}
\end{algorithm}

\subsubsection{Slippage risk} Slippage, the difference between the spot and realized price is a consequence of AMM, and therefore occurs only at AMM DEX. Slippage risk affects only users executing trades - service customers and arbitrageurs - and does not concern LPs. It depends on the size of AMM pool and affects all token pairs. Slippage losses can be amplified by MEV attack.

\subsubsection{Impermanent Loss} Impermanent loss is a consequence of AMM. It affects only LPs to AMM DEX and services customers of yield farming protocols that employ liquidity provisions to AMM DEX in their strategy. Impermanent loss is amplified by crypto-currency volatility and can be avoided by providing liquidity to the token pairs with equal or similar values: \i synthetic tokens targeting the same value (\eg pair of USD stablecoins, SOL wrapped tokens), or \2 synthetic tokens and their target tokens (\eg native tokens and rebase liquid staking tokens).

\color{black}
\subsubsection{Liquidation risk} Liquidation risk affects service customers of DeFi lending protocols - interest rate protocols and crypto-backed stablecoin - and is amplified by crypto-currency volatility. In case of market turmoil with high price volatility, it can also affect LPs when the too-late liquidation of the collateral will not cover the debt position.
\color{black}

\subsubsection{De-peg risk} refers to the circumstances when the synthetic token loses the peg to the reference value. Consequently. It affects all holders of synthetic tokens - stablecoins, wrapped tokens (bridges) and liquid staking tokens. Particularly, it concerns LPs at DEX that provide liquidity for trading synthetic assets.


\section{Discussion}
This section starts with lessons learned from the analysis of DeFi protocols regarding the TVL measure and differentiation between CeFi and DeFi. Further, we present the current trends in DeFi protocol developments and conclude with key challenges that need to be addressed before DeFi gains further adoption.

\subsection{Lessons learned}
 DeFi stakeholders, especially service customers and LPs need to be aware of the counter-party risk related to the DeFi protocols. Clear differentiation between DeFi and CeFi is not straightforward nor clearly communicated by the protocols. The DeFi risks - rug pull, slippage, impermanent loss, liquidation, and de-peg - need to be considered. Moreover, the following aspects need consideration as well.
 





\subsubsection{TVL as a measure of success}
TVL, often used as a success measure in the DeFi, might be misleading for several reasons. First, TVL is a measure of market-making capital locked in DeFi, rather than user activity or traded volume. Second, some protocols, such as DEX aggregators, do not require locking any capital in order to provide services to customers (\eg optimal exchange prices at DEXs). Last, with the emergence of new algorithms \eg concentrated liquidity AMM, higher trading volumes at DEXs can be achieved with less capital locked in AMM pools.

\subsubsection{CeFi vs DeFi}
The framework for classification between CeFi or DeFi\cite{Qin2021CeFiFinance} is not always straightforward to apply. Some protocols, \eg CLOB DEX, fiat-backed stablecoins or tokenized real word assets (RWA), rely both on the blockchain security and integrity as well as on off-chain counterparties. Furthermore, some DeFi protocols, \eg crypto-backed stablecoins rely on the RWA assets and fiat-backed stablecoins, as they are included in the on-chain reserves.

\subsection{Current trends}
DeFi asset management concentrates on optimizing liquidity provision strategies. AMM DEX migrates to CL AMM model for higher capital efficiency. Yield farming protocols look for optimal capital allocation between AMM pools. DeFi vaults, also referred to as indexes are the new category of DeFi protocols that allow participants in the token pool, allocated by the vault manager.

\subsubsection{Concentrated Liquidity AMM} After the introduction of concentrated liquidity AMM, the liquidity provisions strategy at CLAMM DEXs has attracted significantly more TVL compared to other yield farming strategies, \$5.57b and \$1.15b respectively \cite{2022DeFiLlama}. Tokens in a CLAMM DEX are deployed more efficiently, compared to AMM DEX, and allow for higher transaction volume with lover TVL. Examples are the two largest DEXs on Solana: the largest DEX in terms of transaction volume is the second largest DEX in terms of TVL. 

\subsubsection{DeFi Asset Management}
There are many emerging DeFi protocols in the area of asset management. However, their TVL is still low. Two categories worth mentioning included \emph{indexes} and \emph{reserve currencies}. The indexes protocols, such as SetProtocol\cite{2023SetProtocol}, Enzyme\cite{2023Enzyme}, allow DeFi users to create, manage and allocate tokens to vaults, which are similar in their function to the funds and ETF in TradFi. Depositors transfer liquidity into the vault, whereas vault managers allocated tokens to various DeFi protocols to generate yield. Reserve currency protocols function similarly. Depositors transfer tokens into the managed vault. The difference is the valuation of the vault. In the case of the index protocols, the value of the vault is equal to the value of all tokens in the vault. In the case of the reserve currency vault, the value is determined by supply and demand.

\subsection{Future Challenges}
As presented in the risk sections, there are many open challenges hindering the wider adoption of DeFi. Impermanent loss makes the liquidity providing for new tokens unattractive and consequently increase the cost of new token launch. The major obstacle to the lending protocol is over-collateralization which leads to lower capital efficiency and liquidation risk of the collateral.

\subsubsection{Impermanent Loss at AMM DEX}
AMM-based DEX differentiates by utilizing various reserve curves and swap fees, pool weightage, and liquidity mining, mainly in order to mitigate the impermanent loss for the LPs. Liquidity mining refers to the process of collecting native tokens of the DeFi protocol in exchange for participating in the liquidity pool and is a form of subsidizing AMM pools for the token designers. Most DEXs, like Uniswap\cite{Adams2020UniswapCore} or Balancer\cite{MartinelliF2019BalancerSensor.}, offer 50/50 pool weightage, in which LPs must supply the equal value of each token in the liquidity pool. Whereas Balancer\cite{MartinelliF2019BalancerSensor.} offers a variable pool supply criteria. ON Curve\cite{Egorov2021CurvePeg}, the pool weightage is dynamic and does not rebalance its pool. Pool weightage impacts impermanent loss. Further research on the mitigation of impermanent loss by applying other AMM formulas, or using on-chain derivatives, is required. 

\color{black}
Another challenge confronting AMM-based exchanges is the lack of the limit orders that are common in the CLOB exchanges. Only Uniswap (v3) supports range orders that allow replicating some limit order types: take profit orders and buy limit orders. \emph{Range orders} are the special liquidity provision strategy, in which only one token is provided to the liquidity pool, and the price range for the LP position is set outside of the current spot price. Limit order types: buy stop order and stop loss order - can not be replicated using the range orders and are consequently not possible at Uniswap (v3) \cite{2023RangeV3}.
\color{black}

\subsubsection{Over-collateralized Lending}
Although DeFi over-collateralized lending mitigates the risks arising from the anonymous borrowers, it exposes the borrower to the liquidation risk and does not allow to use of the digital asset locked as collateral. Decentralized prime brokerage and on-chain identity - present alternative models for the DeFi undercollateralized lending \cite{BakerLucas2022ParadigmsCredit}.
\emph{Decentralized prime brokerage} protocols - Oxygen\cite{2023Oxygen}, DeltaPrime\cite{2023DeltaPrime}, Gearbox\cite{2023Gearbox} - retain the control over the use of the borrowed funds and identity-based protocols - TrueFi\cite{2023TrueFi}, Goldfinch\cite{2023Goldfinch} provide the mapping between on-chain borrowers and real-world entities.

\subsubsection{Security vs Credit}
Lending protocols compete with the staking process in PoS BC, consequently damaging the network security \cite{Carre2022Security, Chitra2020CompetitiveLicense, Chitra2022ImprovingRedistribution}. When the yield provided by the lending protocols is more attractive than the staking rate (inflation rate), stakers are incentivized to remove their staked tokens and lend them out, thus reducing the network security \cite{Chitra2020CompetitiveLicense}. MEV-redistribution improves the network security by redistributing MEV revenue and disincentivizing to unstake\cite{Chitra2022ImprovingRedistribution}. Some liquid staking protocols, \eg RocketPool\cite{2023RocketPool}, include in their tokens MEV rewards, generated by the validators. The risk of liquid staking includes not only slashing risk, typically for staking, but also de-peg risk, inherent to the synthetic tokens.

\subsubsection{Bridges}

Full collateralization of wrapped tokens is inefficient, but over-collateralization is required due to the volatility of digital assets \cite{Bugnet2022XCC:Assets}. With progress in blockchain interoperability and scalability, multi-chain DEX emerged such as THORChain. They offer swaps of digital assets between BCs.

\subsubsection{Synthetic assets de-peg}
In 2023, the liquid staking protocol Lido\cite{2020Lido:Whitepaper} overtook the decentralized stablecoin protocol MakerDAO \cite{MakerDAOTeam2017TheSystem}, and became the largest DeFi protocol in terms of TVL\cite{2022DeFiLlama}. As indicated in this work, little research has been performed on synthetic assets other than stablecoins. Wrapped tokens and liquid staking tokens are gaining popularity in DeFi, but are exposed to de-peg risk amond other. The stability of their promised peg, especially the inclusion of staking rewards, requires empirical examinations.

\color{black}
\subsubsection{Centralized Finance with On-chain Settlement}
Pursuing definitions of DeFi and CeFi from previous sections, fiat-backed stablecoins and order book exchanges are categorized as CeFi applications with on-chain settlement. Fiat-backed stablecoins maintain the reserves of fiat currencies in the traditional banking system in order to preserve the targeted peg, whereas CLOB exchanges implement order books outside of the blockchain but use the blockchain as a settlement layer. The 0x protocol \cite{Warren20170x:Blockchain} is an example of DEX aggregator that implemented a module with a CLOB exchange for swapping crypto-currencies. The perpetual platform dYdX \cite{Juliano2017DYdX:Derivatives} applied CLOB for trading derivatives (perpetual) on crypto-currencies.
\color{black}

\section{Summary}
This paper analyzes the major categories of DeFi protocols that represent over 85\% of TVL. Both the technical and economical design of DeFi protocols are examined. Our analysis shows that any DeFi protocol can be classified into one of three classes: \1 liquidity pool protocols \2 synthetic (pegged) token protocols or \3 aggregator. Liquidity pool protocols manage the supply and demand sites with smart contracts, \eg interest rate protocols manage supply and demand for loanable funds, and vAMM / AMM DEXs - demand for tradeable tokens and derivatives. Synthetic token protocol issues token with a value pegged to certain targets, \eg stablecoins (peg to the fiat currency or commodity), wrapped tokens (peg to the tokens on other blockchains), liquid staking tokens (peg to the staked tokens, and staking rewards), or index tokens. Aggregators optimize the liquidity provision strategies across DeFi (yield farming protocols) or identify best trade execution at DEX (DEX aggregators). The proposed taxonomy for the chain architecture of DeFi protocols suggests that DeFi cannot be defined as a smart contract-enabled financial application, but a more general definition is required that is based on the security and integrity of blockchains. Finally, we showed that the risk in DeFi depends on three dimensions: \1 DeFi agents, \2 DeFi protocols, and \3 tokens involved. We indicated specifically that service customers and LPs are exposed to different risks, and aggregator protocols accumulate the risk of the protocols and tokens they utilize. 
\color{black}
Lastly, liquid staking tokens, as synthetic (pegged) tokens, because of their design, are prone to the de-peg risk. It should be anticipated that such de-pegs will occur in the future.
\color{black}


\bibliographystyle{IEEEtran}
\bibliography{references}

\end{document}